\documentclass[lettersize,journal]{IEEEtran}
\usepackage{array}
\usepackage[caption=false,font=normalsize,labelfont=sf,textfont=sf]{subfig}
\usepackage{textcomp}
\usepackage{stfloats}
\usepackage{url}
\usepackage{verbatim}
\usepackage{graphicx}
\usepackage{cite}
\usepackage{amsmath,amssymb,amsfonts}
\usepackage{algorithmic}
\usepackage{graphicx}
\usepackage{textcomp}
\usepackage{multirow}
\usepackage{booktabs}
\usepackage{lettrine}
\usepackage{threeparttable}
\usepackage{makecell}
\usepackage{arydshln}
\usepackage{xcolor}
\usepackage{color}
\usepackage[colorlinks=false, citecolor=green, linkcolor=red]{hyperref}
\usepackage[switch]{lineno}

\newcommand{\etal}{\textit{et al}.}
\newcommand{\ie}{\textit{i}.\textit{e}., }
\newcommand{\eg}{\textit{e}.\textit{g}., }

\begin{document}

\title{Self-Supervised Coordinate Projection Network for Sparse-View Computed Tomography}

\author{Qing Wu, \IEEEmembership{Graduate Student Member, IEEE}, Ruimin Feng, Hongjiang Wei, \\ Jingyi Yu, \IEEEmembership{Fellow, IEEE}, and Yuyao Zhang, \IEEEmembership{Member, IEEE}
\thanks{This study was supported by the National Natural Science Foundation of China (No. 62071299, 61901256, 91949120).}
\thanks{Qing Wu is with School of Information Science and Technology, ShanghaiTech University, Shanghai, China, and with Shanghai Advanced Research Institute, Chinese Academy of Sciences, Shanghai, China, and also with University of Chinese Academy of Sciences, Beijing, China (e-mail: wuqing@shanghaitech.edu.cn).}
\thanks{Ruimin Feng and Hongjiang Wei are with School of Biomedical Engineering and Institute of Medical Robotics, Shanghai Jiao Tong University, Shanghai, China (e-mail: \{Fengruimin, hongjiang.wei\}@sjtu.edu.cn).}
\thanks{Jingyi Yu is with School of Information Science and Technology, ShanghaiTech University, Shanghai, China (e-mail: yujingyi@shanghaitech.edu.cn).}
\thanks{Yuyao Zhang (\textit{Corresponding author}) is with School of Information Science and Technology and iHuman Institute, ShanghaiTech University, Shanghai, China (e-mail: zhangyy8@shanghaitech.edu.cn).}
\thanks{This paper has supplementary downloadable material available at http://ieeexplore.ieee.org., provided by the author. The material includes two additional ablation studies for the re-projection strategy in our model.}}

\markboth{Journal of \LaTeX\ Class Files,~Vol.~14, No.~8, August~2021}%
{Wu \MakeLowercase{\textit{et al.}}: Self-Supervised Coordinate Projection Network for Sparse-View CT}

\maketitle
\begin{abstract}
Sparse-view Computed Tomography (SVCT) has great potential for decreasing patient radiation exposure dose during scanning. In this work, we propose a Self-supervised COordinate Projection nEtwork (SCOPE) to reconstruct the artifact-free CT image from the acquired SV sinogram by solving the inverse problem of tomography imaging. To solve the under-determined inverse imaging problem, we first introduce an implicit neural representation (INR) network to constrain the solution space via image continuity prior. And inspired by the relationship between linear algebra and inverse problems, we propose a novel re-projection strategy to generate a dense view sinogram from the initial solution, which significantly improves the rank of the linear equation system and produces a more stable CT image solution space. Specially, the desired CT image is represented as an implicit function of the two-dimensional spatial coordinate to directly approximate the SV sinogram through the CT imaging forward model. Then, a dense-view sinogram is generated from the fine-trained INR network. The final CT reconstruction is reconstructed by applying Filtered Back Projection (FBP) to the generated dense-view sinogram. Additionally, we integrate the recent hash encoding into our SCOPE model, which efficiently accelerates the model training process. We evaluate SCOPE in parallel and fan X-ray beam SVCT reconstruction tasks. Our experiment results demonstrate that the re-projection strategy significantly improves the image reconstruction quality (+3 dB for PSNR at least). The proposed SCOPE model provides state-of-the-art reconstruction results compared to two latest INR-based methods and two well-popular supervised DL methods for the SV CT image reconstruction. The code for this work is available at \url{https://github.com/iwuqing/SCOPE}
\end{abstract} 
\begin{IEEEkeywords}
Sparse-View Computed Tomography, Inverse Imaging Problem, Self-Supervised Learning, Implicit Neural Representation
\end{IEEEkeywords}
\section{Introduction}
\label{sec:introduction}
\par X-ray Computed Tomography (CT) has been widely applied in clinical diagnosis, industrial non-destructive testing, and safety inspection \cite{wang2008outlook,chen2017low}. In recent years, CT played a critical role in the auxiliary diagnosis and disease progress monitoring of COVID-19 pneumonia \cite{long2020diagnosis}. However, the high-level radiation exposure may increase the lifetime risk of cancer, especially for patients undergoing disease monitoring by frequent CT scans such as pneumonia and cancer \cite{hall2008cancer,brenner2007computed}. Therefore, reducing the radiation exposure of CT scans is an urgent need for the current public health status. 
\par Mathematically, the CT acquisition process can be formulated as a linear forward model:
\begin{equation}
    \label{equ-forward-model}
    \mathbf{y} = \mathbf{Ax}+\mathbf{n},
\end{equation}
where $\mathbf{y}\in\mathbb{R}^{N_\mathbf{y}}$ is the measurement data (also known as sinogram), $\mathbf{x}\in\mathbb{R}^{N_\mathbf{x}}$ denotes the CT image to be constructed, $\mathbf{A}\in\mathbb{R}^{N_\mathbf{y}\times N_\mathbf{x}}$ represents the CT forward imaging model (\eg Radon transform operator for parallel X-ray beam CT), and $\mathbf{n}\in\mathbb{R}^{N_\mathbf{y}}$ is the system noise. To reduce the imaging radiation dose, one can decrease the dimension of measurement data, denoting as $\mathbf{y_{s}}$, an undersampling of sinogram $\mathbf{y}$. To reconstruct a CT image from the under-sampled sinogram $\mathbf{y_{s}}$ is referred to as Sparse-View (SV) CT reconstruction, a highly ill-posed inverse problem.Analytical reconstruction algorithms, such as Filtered Back-Projection (FBP) \cite{fbp}, results in severe streaking artifacts on the constructed CT image \cite{frikel2013characterization, storath2015joint}.
\begin{table*}[t]
\centering
\caption{Comparison of the proposed SCOPE with existing INR-based SVCT reconstruction methods.}
\label{table-inr-comparison}
\resizebox{\textwidth}{!}{
\begin{tabular}{llllrrr} 
\toprule
\textbf{Methods}    & \textbf{Key Factors for SVCT}    & \textbf{Input} & \textbf{Output}& \textbf{Learning Target}& \textbf{Main Objective}& \textbf{Encoding Strategy} \\ 
\midrule
\textbf{GRFF} \cite{tancik2020fourier}& INR Implicit Prior                    & SV sinogram $\mathbf{y}_s$                    & CT image $\hat{\mathbf{x}}$&CT image $f_\mathbf{\Theta}: (x, y) \rightarrow \mathbf{x}$&   ${\arg \min}_\mathbf{\Theta} \,\mathcal{L}(\mathbf{A}f_\mathbf{\Theta}, \mathbf{y}_s)$      &  Fourier encoding       \\\midrule
\textbf{IntroTomo} \cite{zang2021intratomo} & \makecell[l]{INR Implicit Prior +\\ Explicit Prior}                    & SV sinogram $\mathbf{y}_s$                    & CT image $\hat{\mathbf{x}}$                 &CT image $f_\mathbf{\Theta}: (x, y) \rightarrow \mathbf{x}$& ${\arg \min}_\mathbf{\Theta} \,\mathcal{L}(\mathbf{A}f_\mathbf{\Theta}, \mathbf{y}_s)$  &Fourier encoding\\\midrule
\textbf{NeRP} \cite{shen2021nerp}& \makecell[l]{INR Implicit Prior +\\ Longitudinal Image Prior}                    & \makecell[l]{A prior CT image + \\ SV sinogram $\mathbf{y}_s$}                    & CT image $\hat{\mathbf{x}}$                 &CT image $f_\mathbf{\Theta}: (x, y) \rightarrow \mathbf{x}$&${\arg \min}_\mathbf{\Theta} \,\mathcal{L}(\mathbf{A}f_\mathbf{\Theta}, \mathbf{y}_s)$  & Fourier encoding\\\midrule
\textbf{CoIL} \cite{sun2021coil}& DV Sinogram Generation                    & SV sinogram $\mathbf{y}_s$                    & DV sinogram $\hat{\mathbf{y}}_d$   & Sinogram $g_\mathbf{\Theta}: (\theta, \rho) \rightarrow \mathbf{y}$  &  ${\arg \min}_\mathbf{\Theta} \,\mathcal{L}(g_\mathbf{\Theta}, \mathbf{y}_s)$  & Linear encoding      \\\midrule
\textbf{SCOPE (Ours)}& \makecell[l]{INR Implicit Prior + \\ DV Sinogram Generation}                     & SV sinogram $\mathbf{y}_s$                    & DV sinogram $\hat{\mathbf{y}}_d$                 &CT image $f_\mathbf{\Theta}: (x, y) \rightarrow \mathbf{x}$&${\arg \min}_\mathbf{\Theta} \,\mathcal{L}(\mathbf{A}f_\mathbf{\Theta}, \mathbf{y}_s)$  & Hash encoding\\
\bottomrule
\end{tabular}}
\end{table*}
\par Many attempts have been made to eliminate the streaking artifacts on the restructured CT images. Conventional iterative reconstruction methods \cite{sidky2006accurate,kim2014sparse,sidky2008image,rudin1992nonlinear} formulate this under-determined inverse imaging as a regularized optimization problem. Explicit image prior assumptions (\eg Total Variation (TV) \cite{rudin1992nonlinear} for inducing smoothness in CT image) are adopted as a regularization term to provide an approximate solution \cite{liu2021zero}. Recently, supervised Deep Learning (DL) methods \cite{FBPConvNet,han2018framing,zhang2018sparse,lee2018deep,shen2019r,li2019learning,ding2021learnable} have shown great potential for SVCT reconstruction. Instead of directly solving this ill-posed inverse problem, a supervised DL reconstruction mostly employs Convolutional Neural Network (CNN) to learn an end-to-end mapping from reconstructed low-quality CT images to the corresponding high-quality images over a large dataset. For example, {Jin \etal} \cite{FBPConvNet} proposed FBPConvNet that trains a U-Net \cite{Unet} to learn the residual from artifact-corrupted inputs to artifact-free outputs. It is known that the performance of the supervised DL methods highly depends on the data distribution of the image pairs in the training dataset (\ie a large-scale training dataset that covers more types of variations generally provides better performance). However, the generalization issue is that the hyperparameters would differ if the training SVCT images are different. Differences in different training datasets,  such as SV undersampling schemes, beam types for measurement data projection, different organs, would significantly affect the performance of the trained networks. 

\par {The} Implicit Neural Representation (INR) has recently been proposed to model and represent 3D scenes from a sparse set of 2D views using coordinate-based deep neural networks in a self-supervised fashion. The core component in INR is a continuous implicit function parameterized by a Multi-Layer Perceptron (MLP). Benefiting from the image continuity prior imposed by the implicit function and the neural network architecture, INR has achieved superior performance in various vision problems (\eg surface reconstruction \cite{park2019deepsdf, chen2019learning, mescheder2019occupancy}, view synthesis \cite{mildenhall2020nerf,zhang2020nerf++, rebain2021derf}, and image super-resolution \cite{chen2021learning,tang2021joint}).

\par For SVCT imaging, an early attempt was made by Tancik \etal \cite{tancik2020fourier} indicated that INR could be applied to recover the CT image from the collected SV sinogram without using any external data. Since then, a few INR-based works \cite{sun2021coil, shen2021nerp, zang2021intratomo, reed2021dynamic, vasconcelos2022uncertainr, gupta2022differentiable} have emerged for the reconstruction of CT images. We summarize the recent works that solve the inverse problem of  tomography imaging using INR-based methods in Table \ref{table-inr-comparison}, to compare the design ideas and characteristics of various methods more clearly. Sun \etal \cite{sun2021coil} proposed CoIL that trains an INR to represent the SV sinogram and predicts the accordance Dense-View (DV) sinogram based on the continuous nature of INR. The CT image reconstruction is then processed by applying user-chosen reconstruction methods (e.g., FBP \cite{fbp}) on the predicted DV sinogram. However, the coordinate space of the sinogram does not follow the intuitive orthogonal assumption of Fourier spatial encoding in the INR model. Thus the performance of CoIL for CT reconstruction is not comparable with supervised DL methods. Shen \etal \cite{shen2021nerp} proposed NeRP to utilize a series of longitudinal CT scans of the same subject to build CT image from SV sinogram. The INR is firstly trained on a high-quality DV CT scan then used as an image prior to generate the high-quality CT images from the acquired SV sinograms. However, longitudinal CT scans from the same patient are not always available. Zang \etal \cite{zang2021intratomo} proposed IntroTomo that combines a sinogram prediction module with a geometry refinement module. The former module used INR to reconstruct the CT image from the SV sinogram, while the latter module combines explicit priors (TV and {non-local mean}) via an optimization framework to refine CT images. The two modules are trained iteratively to improve the CT image quality but severely prolong reconstruction time. 
\par Compared with the works in Table \ref{table-inr-comparison}, our proposed method is most related to in that {Refs. \cite{tancik2020fourier, sun2021coil}}. However, there are two major limitations unsolved in those works: 1) The INR estimated the desired CT image by minimizing the loss between the network-predicted sinogram and the acquired sinogram. Thus the paradigm is more efficient in sinogram generation rather in CT image reconstruction. Due to the highly sparse sinogram, the MLP tends to approach an implicit function that may overfit the SV sinogram, which manifests as noisy INR-represented CT images; 2) Due to the heavy computation of the coordinate-based based deep MLP, the image-specific INR-based CT reconstructions generally performs poorly on time-efficiency.
\par In this paper, we propose a Self-supervised COordinate Projection nEtwork (SCOPE) to reconstruct the high-quality artifact-free CT image from the acquired SV sinogram by solving the ill-posed inverse problem of tomography imaging without any external data. We first introduce an implicit neural representation (INR) network to constrain the solution space via image continuity prior to achieve an initial solution of the CT image. Compared with existing related works \cite{tancik2020fourier, zang2021intratomo, shen2021nerp, sun2021coil}, one of our key contributions is a simple and effsective re-projection strategy that significantly improves the quality of reconstructed CT images. This strategy is inspired by the relationship between linear algebra and inverse problems. We consider the SVCT inverse imaging problem as an under-determined system of linear equations. The total number of X-rays involved in all sinograms is equivalent to the number of independent linear equations (\ie the rank of a matrix $\mathbf{A}$ in Eq. \ref{equ-forward-model}). Thus the number of free variables in the linear equations largely increases with the decrease of the matrix $\mathbf{A}$’s rank in the SV sinogram. By introducing {the} INR, the solution space of image $\mathbf{x}$ is efficiently constrained in a continuous space, resulting a satisfied inverse CT reconstruction from a highly sparse sinogram. However, the reconstructed signal intensity of the CT image is ambiguous, which can be easily affected by network overfitting to the SV sinogram. Here, we propose a novel re-projection strategy to build a DV sinogram from this initial CT reconstruction. {Benefiting from the continuous} nature of the INR represented CT image, this process is equivalent to generating a high-rank linear equation system. Our experiment results demonstrate that through this re-projection strategy, the image noise is further suppressed with preserved image details in the reconstructed CT images, resulting in an improved image quality (+3 dB for PSNR at least). In addition, learning high-frequency signals via simple MLP is practically very difficult due to the spectral bias problem \cite{rahaman2019spectral, xu2019frequency}. Existing INR-based methods have low reconstruction efficiency since they mostly use deep MLPs with pre-defined encoding modules (\eg Fourier encoding \cite{shen2021nerp}) to learn the implicit function. To accelerate the model training, we integrate the recent hash encoding \cite{muller2022instant} into our SCOPE model, enabling shallow (three-layers) MLP achieve superior fitting ability (about 1 minute). We evaluated our proposed method on two public datasets (AAPM and COVID-19). Both qualitative and quantitative results indicate that SCOPE provides state-of-the-art reconstruction results compared to two recent INR-based methods (CoIL \cite{sun2021coil} and GRFF \cite{tancik2020fourier}) and two well-known supervised CNN-based models (FBPConvNet \cite{FBPConvNet} and TF U-Net \cite{han2018framing}). To the best of our knowledge, the proposed SCOPE is the first self-supervised method that outperforms the supervised DL models for SVCT reconstruction. The main contributions of this work are summarized as below:
\begin{enumerate}
    \item We propose SCOPE that recovers the high-quality CT image from acquired SV sinogram without involving any external data.
    \item We propose a simple and effective re-projection reconstruction strategy that significantly improves the quality of reconstructed CT images.
    \item We integrate the hash encoding \cite{muller2022instant} into our SCOPE, which greatly accelerates the model training and thus improves the model practicability.
\end{enumerate}
\section{Methodology}
\label{sec:methodology}
\subsection{Overview}
\par In the proposed SCOPE model, we represent the desired CT image $\mathbf{x}$ as a continuous function parameterized by a neural network:
\begin{linenomath*}
\begin{equation}
    I = f_\mathbf{\Theta}(\mathbf{p}),
\end{equation}
\end{linenomath*}
where $\mathbf{\Theta}$ denote the trainable parameters (weights and biases) of the network, $\mathbf{p}=(x,y)\in \mathbb{R}^2$ is any 2D spatial coordinate in the imaging plane, and $I\in \mathbb{R}$ is the corresponding image intensity at the position $\mathbf{p}$ in the image $\mathbf{x}$. Based on the acquired SV sinogram $\mathbf{y}_s$, we then optimize the network to approximate the implicit function using a back-propagation gradient descent algorithm to minimize the objective as below:
\begin{linenomath*}
\begin{equation}
    \hat{\mathbf{\Theta}}=\underset{\mathbf{\Theta}}{\arg \min} \, \mathcal{L}(\hat{\mathbf{y}}_s, \mathbf{y}_s),\ \text{with}\ \hat{\mathbf{y}}_s=\mathbf{A}f_\mathbf{\Theta},
    \label{equ:obj}
\end{equation}
\end{linenomath*}
where $\hat{\mathbf{y}}_s$ represents the predicted SV sinogram and $\mathcal{L}$ is the loss function that measures the discrepancy between the predicted SV sinogram $\hat{\mathbf{y}}_s$ and the acquired SV sinogram $\mathbf{y}_s$. 
\par The key insight behind Eq. \ref{equ:obj} is using the image continuity prior imposed by the implicit function and the neural network architecture to regularize the inverse imaging problem of SVCT and thus obtaining the desired solution. After the network training, the optimal image $\hat{\mathbf{x}}$ is theoretically $f_{\hat{\mathbf{\Theta}}}$. However, due to the highly under-determined inverse imaging problem, the network tends to approach an implicit function that overfits the SV sinogram $\mathbf{y}_s$ and thus fails to approximate the desired implicit function well, which {manifests} as severe noise on the reconstructed CT image $\hat{\mathbf{x}}=f_{\hat{\mathbf{\Theta}}}$.
\par
\begin{figure}[t]
    \centering
    \includegraphics[width=0.95\linewidth]{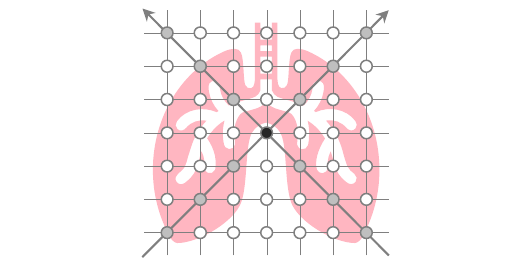}
    \caption{A toy example of different types of sample points in SVCT: \textbf{Black} sample points are scanned by multiple X-rays, whose pixel intensities are well constrained in the inverse imaging problem; \textbf{Gray} sample points are scanned by a single X-ray; \textbf{White} sample points are not scanned by any X-ray. The gray and white points are examples of free variable pixels, whose intensities are not tightly constrained in the inverse problem.}
    \label{fig:scan_points}
\end{figure}

\begin{figure*}[t]
    \centering
    \includegraphics[width=0.95\textwidth]{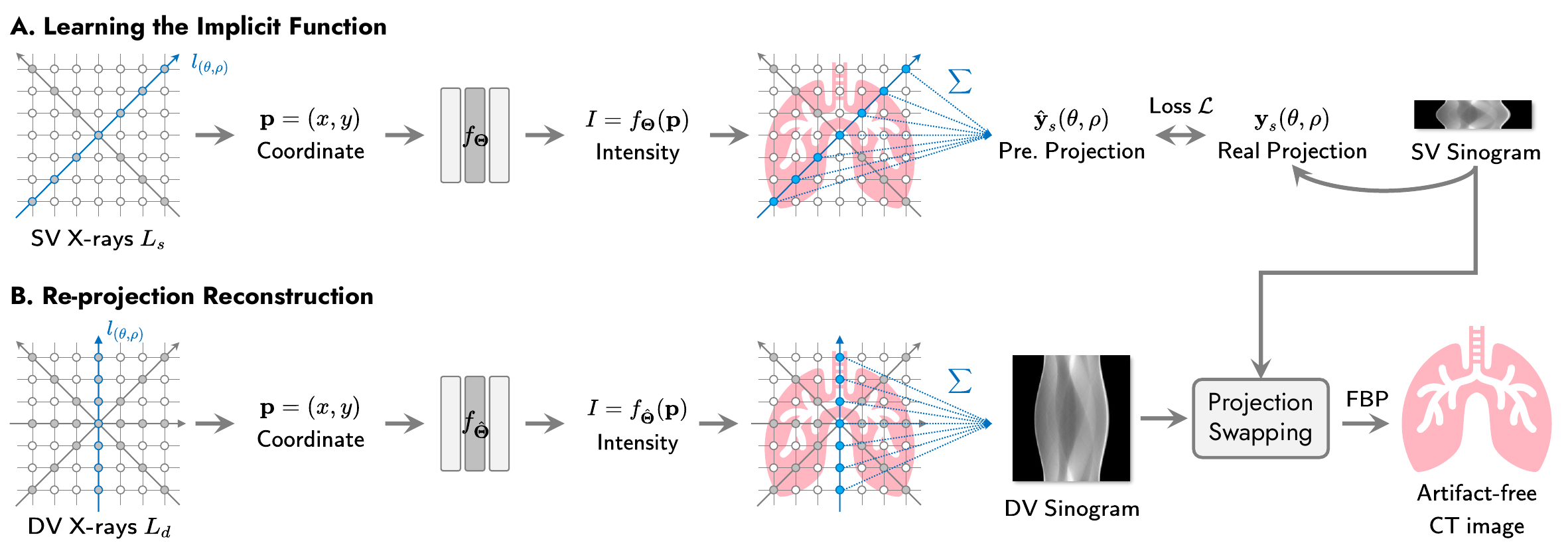}
    \caption{Workflow of the proposed SCOPE model. \textbf{A. Learning the Implicit Function}: The network parameterizing implicit function $f_{\mathbf{\Theta}}$ takes the coordinate $\mathbf{p}$ of sampling points along SV X-rays $L_s$ as input and predicts the image intensity $I= f_{\mathbf{\Theta}}(\mathbf{p})$ at these positions. Then, the projections $\hat{\mathbf{y}_s}(\theta, \rho)$ of the X-rays $L_s$ are calculated by the summation operator (Eq. \ref{equ_projection_dis}). Finally, we optimize the network by minimizing the loss between the predicted projections $\hat{\mathbf{y}_s}(\theta, \rho)$ and real projection $\mathbf{y}_s(\theta, \rho)$ from acquired SV sinogram. \textbf{B. Re-projection Reconstruction}: The coordinates $\mathbf{p}$ of sample points along DV X-rays $L_d$ are fed into the well-trained network to estimate the corresponding image intensity $I= f_{\hat{\mathbf{\Theta}}}(\mathbf{p})$. Similarly, the projections $\hat{\mathbf{y}_d}(\theta, \rho)$ of the X-rays $L_d$ are computed by the summation operator. {Next, we apply a projection swapping operator to combine the generated projections with the actual SV sinogram, producing the final DV sinogram. The CT image is then reconstructed using FBP \cite{fbp} applied to the DV sinogram.}}
    \label{fig:pipline}
\end{figure*}

 To this end, we propose a re-projection reconstruction strategy, in which the learned function $f_{\hat{\mathbf{\Theta}}}$ is used to generate a DV sinogram $\hat{\mathbf{y}}_d$. Then the final high-quality CT image $\hat{\mathbf{x}}$ is reconstructed by applying FBP \cite{fbp} on $\hat{\mathbf{y}}_d$. An essential insight is that the INR network overfitting on the SV sinogram results in unexpected pixel intensity mutations in the CT image reconstruction. Fig. \ref{fig:scan_points} illustrates a toy example of different types of sample points in SV reconstructed CT. For example, the black sample points are scanned by multiple X-rays, which can be considered as constrained by multiple linear equations. Thus the INR network can accurately recover its image intensity through the constraints of the cross projections. For the gray and white sample points scanned only by few, or even no X-rays, the pixel intensities are not tightly constrained in the inverse problem. These pixel intensities are mostly approximated by the image continuity prior imposed by the implicit function and are easily affected by the overfitting effected towards the sparse measurements of the sinogram. Therefore, the learned function $f_{\hat{\mathbf{\Theta}}}$ may output pixel intensity mutation at those free variable positions due to the overfitting problem. Although these mutations manifest similarly to image noise, they do not follow any typical distribution, thus the performance of inserting common denoising regularization term is limited \cite{ zang2021intratomo}. The most effective strategy to suppress free variable mutations is thus to generate a higher-rank linear equation system that tightly constrains the pixel intensities in the CT image and produces the same solution space with the SV sinogram. The generation of a DV sinogram $\hat{\mathbf{y}}_d$ from  $f_{\hat{\mathbf{\Theta}}}$ is thus proposed. The workflow of the proposed SCOPE model is shown in Fig. \ref{fig:pipline}.
\subsection{Learning {the} Implicit Function}
\par Fig. \ref{fig:pipline} \textbf{A} demonstrates the pipeline of learning implicit function by a neural network. Given a SV sinogram $\mathbf{y}_s\in \mathbb{R}^{K\times M}$, where $K$ and $M$ are the number of projection views and X-rays per view respectively, we first build a total number of $K*M$ X-rays $L_s$ from the $K$ sparse projection views (\ie $M$ X-rays per view). Next, we feed the spatial coordinates $\mathbf{p}$ of sample points along the SV X-rays $L_s$ into the implicit function to produce the corresponding image intensities $I=f_\mathbf{\Theta}(\mathbf{p})$. Finally, we compute the predicted projection $\hat{\mathbf{y}}_s(\theta,\rho)$ of each one $l_{
(\theta, \rho)}:y\sin\theta+x\cos\theta=\rho$ in the X-rays $L_s$ by a summation operator as below:
\begin{linenomath*}
\begin{equation}
    \hat{\mathbf{y}}_s(\theta,\rho) = \sum_{\mathbf{p}\in l_{(\theta, \rho)}}f_\mathbf{\Theta}(\mathbf{p}),
    \label{equ_projection_dis}
\end{equation}
\end{linenomath*}
where $\theta=\{\theta_i\}_{i=1}^{K}$ are the sparse projection views and $\rho=\{\rho_j\}_{j=1}^{M}$ are the positions of X-rays in the detector. 
\par Since the summation operator (Eq. \ref{equ_projection_dis}) is differentiable, the neural network used for parameterizing the implicit function $f_\mathbf{\Theta}$ can be optimized by using back-propagation gradient decent algorithm to minimize the loss between the predicted projection $\hat{\mathbf{y}}_s(\theta,\rho)$ and the real projection $\mathbf{y}_s(\theta,\rho)$ from the SV sinogram $\mathbf{y}_s$. In this work, we employ $\ell_1$ norm as the loss function, which is defined as below:
\begin{linenomath*}
\begin{equation}
    \mathcal{L}=\frac{1}{k*m}\sum_{i=1}^{k} \sum_{j=1}^{m} \left| \mathbf{y}_s(\theta_i, \rho_j) - \hat{\mathbf{y}}_s(\theta_i, \rho_j)\right|,
    \label{Eq.loss}
\end{equation}
\end{linenomath*}
where $k$ and $m$ are respectively the number of sampled projection views and the sampled X-rays per view at each training iteration.
\subsection{Re-projection Reconstruction}
\label{sec:Re-projection Reconstruction}
\par
\begin{figure}[t]
    \centering
    \includegraphics[width=0.9\linewidth]{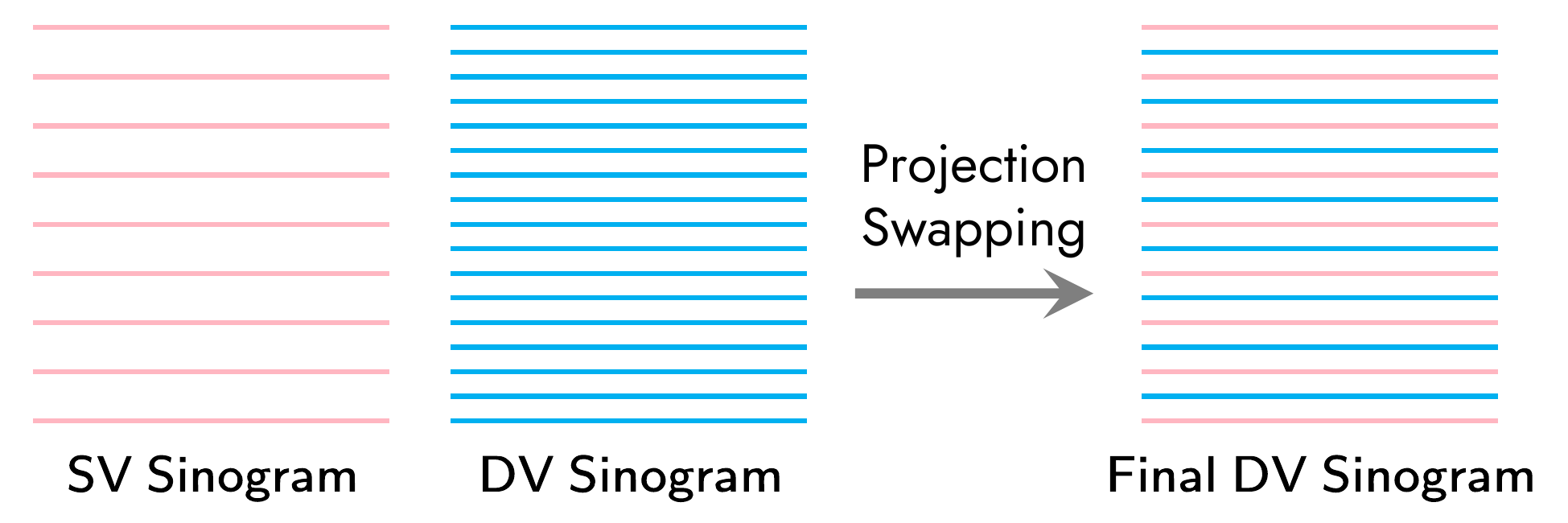}
    \caption{{Pipeline of the projection swapping operator.}}
    \label{fig:projection swapping}
\end{figure}

 Fig. \ref{fig:pipline} \textbf{B} shows the workflow of the proposed re-projection reconstruction strategy, in which the learned implicit function $f_{\hat{\mathbf{\Theta}}}$ is used to generate the DV sinogram $\hat{\mathbf{y}}_d\in \mathbb{R}^{K_d\times M}$ then the final high-quality CT image $\hat{\mathbf{x}}$ is reconstructed from the DV sinogram. More specifically, we first build $K_d*M$ X-rays $L_d$ from $K_d$ dense projection views (\ie $M$ X-rays per view). Then, the spatial coordinates $\mathbf{p}$ of the sample points along the DV X-rays $L_d$ are fed into the learned function to predict the corresponding image intensities $I=f_{\hat{\mathbf{\Theta}}}\mathbf{p}$. Similarly, the projection $\hat{\mathbf{y}}_d$ of the X-rays $L_d$ are also calculated by the summation operator (Eq. \ref{equ_projection_dis}). The DV sinogram $\hat{\mathbf{y}}_d$ is thus generated. Inspired by the data consistency used in MRI acceleration reconstruction \cite{yaman2020self}, we combine the estimated DV sinogram $\hat{\mathbf{y}}_d$ with the acquired SV sinogram $\mathbf{y}_s$ to generate the final DV sinogram, {as shown in Fig. \ref{fig:projection swapping}}. In particular, we replace the projection profiles at the corresponding views in the DV sinogram $\hat{\mathbf{y}}_d$ with the acquired SV sinogram $\mathbf{y}_s$. {The effectiveness of the projection swapping operator is discussed in the supplementary material.} Finally, we apply FBP \cite{fbp} on the final DV sinogram to reconstruct the artifact-free CT image.
\subsection{Network Architecture}
\par
\begin{figure}[t]
    \centering
    \includegraphics[width=0.9\linewidth]{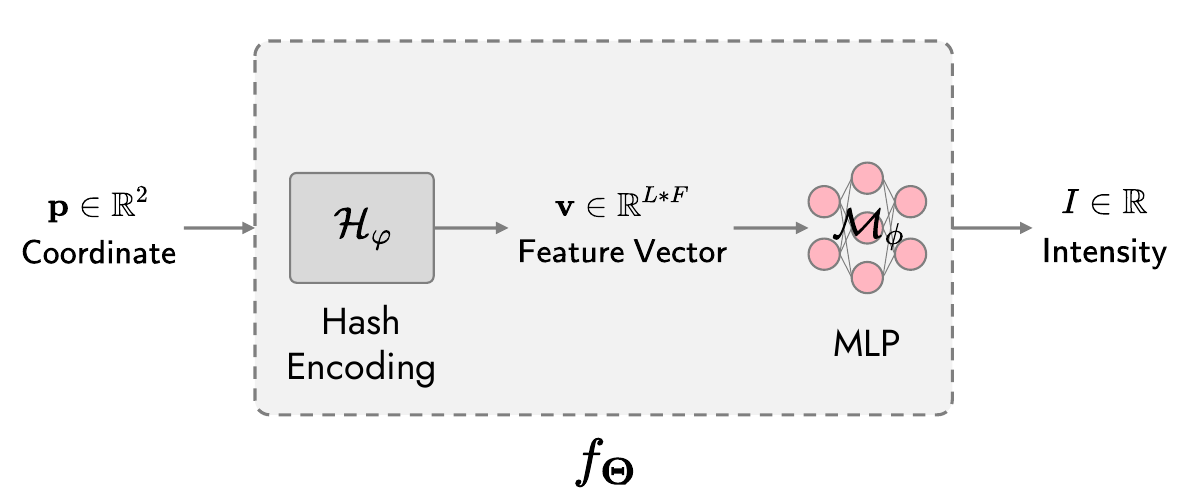}
    \caption{The architecture of the neural network used for parameterizing the implicit function $f_\mathbf{\Theta}$, which consists of the hash encoding \cite{muller2022instant} and a three-layers MLP.}
    \label{fig:network_architecture}
\end{figure}

 As shown in Fig. \ref{fig:network_architecture}, the network used for learning the implicit function $f_\mathbf{\Theta}$ consists of an encoding module (via hash encoding \cite{muller2022instant}) and a three-layers MLP. The network maps the input coordinate $\mathbf{p}$ to a feature vector $\mathbf{v}\in \mathbb{R}^{L*F}$ and then converts the feature vector $\mathbf{v}$ to the image intensity $I$. This process can be expressed as below:
\begin{linenomath*}
    \begin{equation}
        I=\mathcal{M}_\phi(\mathbf{v}),\quad \mathbf{v}=\mathcal{H}_{\varphi}(\mathbf{p}),
    \end{equation}
\end{linenomath*}
where $\phi$ and $\varphi$ represent respectively the trainable parameters of the MLP and hash encoding. They are simultaneously optimized to estimate the implicit function $f_\mathbf{\Theta}$.
\subsubsection{Hash Encoding} The universal approximation theorem \cite{hornik1989multilayer} proved that a pure MLP could approximate any complicated function theoretically. However, fitting high-frequency signals via the pure MLP is practically very difficult due to the spectral bias problem \cite{rahaman2019spectral,xu2019frequency}. To alleviate the issue, many encoding strategies \cite{muller2022instant,sun2021coil,tancik2020fourier,mildenhall2020nerf} have been proposed to map low-dimensional inputs into high-dimensional feature vectors, which allows the subsequent MLP to capture high-frequency components easily and thus reduce approximation error. In SCOPE, we adopt recent hash encoding. Unlike pre-defined encoding rules (\eg position encoding \cite{mildenhall2020nerf}), hash encoding assigns a trainable feature for each input coordinate. This adaptive encoding strategy is task-specific, which benefit from using a shallow MLP while achieving powerful fitting ability. For a coordinate grid of $N\times N$, hash encoding first builds multi-resolution of $L$ levels feature maps $\{\mathbf{V}_i\}_{i=1}^{L}$. Here $\mathbf{V}_i\in \mathbb{R}^{N_i\times N_i\times F}$ is the feature map at the $i$-th level, where each element is a trainable feature vector of $F$ length. Then, each feature map $\mathbf{V}_i$ is mapped into a hash table of $T$ size to reduce memory footprint. After the hash table construction, given input coordinate $\mathbf{p}$, we compute its feature vector $\mathbf{v}_i\in \mathbb{R}^{F}$ at the $i$-th level via trilinear interpolation. Then, we concatenate $L$ feature vectors $\{\mathbf{v}_i\}_{i=1}^{L}$ to produce the final feature vector $\mathbf{v}\in\mathbb{R}^{L*F}$. In our SCOPE model, the hyper-parameters of the hash encoding are set as follow: $L=8$, $T=2^{24}$, $F=8$, $N_\text{min}=2$, and $b=2$. More details about the hash encoding can be found in {Ref.} \cite{muller2022instant}. 
\subsubsection{Three-Layers MLP} After the hash encoding, the 2D input coordinate $\mathbf{p}\in\mathbb{R}^{2}$ is encoded to the high-dimensional feature vector $\mathbf{v}\in\mathbb{R}^{L*F}$. Then, a three-layers MLP is used to convert the feature vector $\mathbf{v}$ to the image intensity $I$. The two hidden layers in the MLP have 64 neurons and are followed by ReLU activation function and the output layer is followed by Sigmoid activation function.
\subsection{Training Parameters}
\par For the training of the proposed SCOPE model, at each iteration, we first randomly sample 3 ones (\ie $k=3$ in Eq. \ref{Eq.loss}) from sparse projection views $\{\theta_i\}_{i=1}^K$ and then randomly sample 10 ones (\ie $m=10$ in Eq. \ref{Eq.loss}) from $M$ X-rays per view. We adopt Adam optimizer \cite{Kingma2015AdamAM} to minimize the $\ell_1$ loss function and the hyper-parameters of the Adam are as follows: $\beta_1=0.9,\ \beta_2=0.999,\ \epsilon=10^{-8}$. The initial learning rate is $10^{-3}$ and decays by a factor of 0.5 per 500 epochs. The total number of training epochs is 5000, which only takes about 5 minutes on a single NVIDIA RTX 3060 GPU. It is worth noting that all the training parameters above are the same for different cases, such as different types of X-ray beam and input views.

\section{Experiments}
\label{sec:exp}
\begin{table}[t]
\centering
\caption{Hyper-parameters of the four built-in functions in MATLAB R2021b used for data simulation.}
\label{table-data-simulation}
\resizebox{0.415\textwidth}{!}{
\begin{threeparttable}
\begin{tabular}{llr} 
\toprule
\textbf{Function}        & \textbf{Hyper-parameter} & \textbf{Value}  \\ 
\midrule
\texttt{radon}                     & theta                    & $\{(i-1)\times 180/k\}^{k}_{i=1}$                 \\
\midrule
\multirow{2}{*}{\texttt{iradon}}   & theta                    & $\{(i-1)\times 180/k\}^{k}_{i=1}$                 \\
                          & output\_size             & $h\times w$                 \\
\midrule
\multirow{3}{*}{\texttt{fanbeam}}  & D                        &$\sqrt{h^2 + w^2}$                  \\
                          & FanRotationIncrement     &$360/k$                  \\
                          & FanSensorSpacing         &$0.1$                  \\
\midrule
\multirow{4}{*}{\texttt{ifanbeam}} & D                        &$\sqrt{h^2 + w^2}$                  \\
                          & FanRotationIncrement     &$360/k$                  \\
                          & FanSensorSpacing         &$0.1$                  \\
                          & OutputSize~              &$h\times w$                  \\
\bottomrule
\end{tabular}
\begin{tablenotes}
\footnotesize
\item[$\star$] $k$ is the number of projection views and $h\times w$ are the size of raw slice.
\end{tablenotes}
\end{threeparttable}}
\end{table}
\subsection{Dataset \& Pre-processing}
\subsubsection{AAPM dataset} Based on the normal dose part of the 2016 low-dose CT challenge AAPM dataset\footnote{https://www.aapm.org/GrandChallenge/LowDoseCT/} that consists of twelve 3D CT volumes acquired from twelve subjects, the AAPM dataset used in our experiments is built. Specifically, we extract 1171 2D slices from the 3D CT volumes on axial view and then split these slices into three parts: 1069 slices from ten subjects in training set, 98 slices from one subject in validation set, and 4 slices from one subject in test set. \textit{The training and validation sets are only prepared for optimizing two supervised CNN-based baselines} (FBPConvNet \cite{FBPConvNet} and TF U-Net \cite{han2018framing}), while other methods (FBP \cite{fbp}, CoIL \cite{sun2021coil}, GRFF \cite{tancik2020fourier}, and our SCOPE) directly recover the corresponding high-quality CT image from the single SV sinogram. 
\subsubsection{COVID-19 dataset} COVID-19 dataset \cite{shakouri2021covid19} is a large-scale CT dataset, which consists of 3D CT volumes from 1000+ patients with confirmed COVID-19 infections. A 3D CT volume of the COVID-19 dataset is employed as additional test data. We select 4 slices from the volume on axial view as 4 test samples. 
\subsubsection{Dataset Simulation} For the parallel and fan X-ray beam SVCT reconstruction, we follow the strategies in \cite{FBPConvNet,han2018framing,shen2019r} to simulate the pairs of low-quality and high-quality CT images. Specifically, we first generate the sinograms of different views (720, 120, 90, and 60) by projecting the raw slices using the built-in functions \texttt{radon} and \texttt{fanbeam} in MATLAB R2021b, respectively. Then, we transfer the sinograms back to CT images using the built-in functions \texttt{iradon} and \texttt{ifanbeam} in MATLAB R2021b, respectively. Detailed hyper-parameters of the four functions are demonstrated in Table \ref{table-data-simulation}. The images reconstructed from 720 views are used for Ground Truth (GT), while the images reconstructed from 120, 90, and 60 views are used for input images corresponding to three different factors $6\times$, $8\times$, and $12\times$. Note that the parallel and fan X-ray beam SVCT are considered as two independent reconstruction tasks. Thus, all the training and test processes are solely conducted.
\subsection{Compared Methods \& Evaluation Metrics}
\subsubsection{Compared Methods} We compare the proposed SCOPE model with five SVCT reconstruction methods: 1) FBP \cite{fbp}, a classical analytical reconstruction algorithm; 2) CoIL \cite{sun2021coil}, an INR-based method. Since the output of CoIL is the DV sinogram. we thus apply FBP on the generated DV sinogram to reconstruct the CT image; 3) GRFF \cite{tancik2020fourier}, an INR-based method with Gaussian random Fourier feature encoding strategy; 4) FBPConvNet \cite{FBPConvNet}, a supervised DL method based on U-Net \cite{Unet}; 5) TF U-Net \cite{han2018framing}, a supervised DL method based on Tigh Frame U-Net. We train FBPConvNet and TF U-Net on the training set of the AAPM dataset through Adam optimizer \cite{Kingma2015AdamAM} with a mini-batch of 8. The learning rate starts from 10$^{-3}$ to 10$^{-6}$, which gradually decreases over each training epoch. The total training epochs are set as 500 and the best model is saved by checkpoints during the training process. The two INR-based methods (CoIL and GRFF) are implemented following the original papers.
\begin{table}[t]
\centering
\caption{Quantitative results (PSNR/SSIM) of the DV sinograms generated by different methods on the COVID-19 dataset for \textit{parallel} and \textit{fan} X-rays beam SVCT of 60, 90, and 120 views. The best performances are highlighted in bold.}
\label{table:ablation-sinogram}
\resizebox{0.93\linewidth}{!}{
\begin{tabular}{clccc}
\toprule
\textbf{X-ray}                & \textbf{Method} & \textbf{60 Views} & \textbf{90 Views} & \textbf{120 Views}  \\ 
\midrule
\multirow{4}{*}{\textit{Parallel}} & Linear            & $34.64/0.9502$      & $37.83/0.9723$      & $40.33/0.9832$        \\
                                   & Cubic          & $34.38/0.9470$      & $37.59/0.9710$      & $40.13/0.9826$        \\
                                   & CoIL \cite{sun2021coil}          & $47.29/0.9915$      & $51.94/0.9968$      & $55.00/0.9985$        \\ 
                                   & SCOPE          & $\mathbf{53.27/0.9977}$      & $\mathbf{58.42/0.9993}$      & $\mathbf{60.60/0.9995}$        \\ 
\midrule
\multirow{4}{*}{\textit{Fan}} & Linear            & $30.86/0.9124$      & $33.96/0.9426$      & $36.38/0.9613$        \\
                                   & Cubic          & $30.61/0.9046$      & $33.75/0.9381$      & $36.20/0.9588$        \\
                                   & CoIL \cite{sun2021coil}          & $40.83/0.9734$      & $45.30/0.9869$      & $48.66/0.993$        \\ 
                                   & SCOPE          & $\mathbf{56.52/0.9989}$      & $\mathbf{59.86/0.9995}$      & $\mathbf{61.37/0.9996}$        \\ 
\bottomrule
\end{tabular}}
\end{table}
\begin{figure}[t]
    \centering
     {\includegraphics[width=0.48\textwidth]{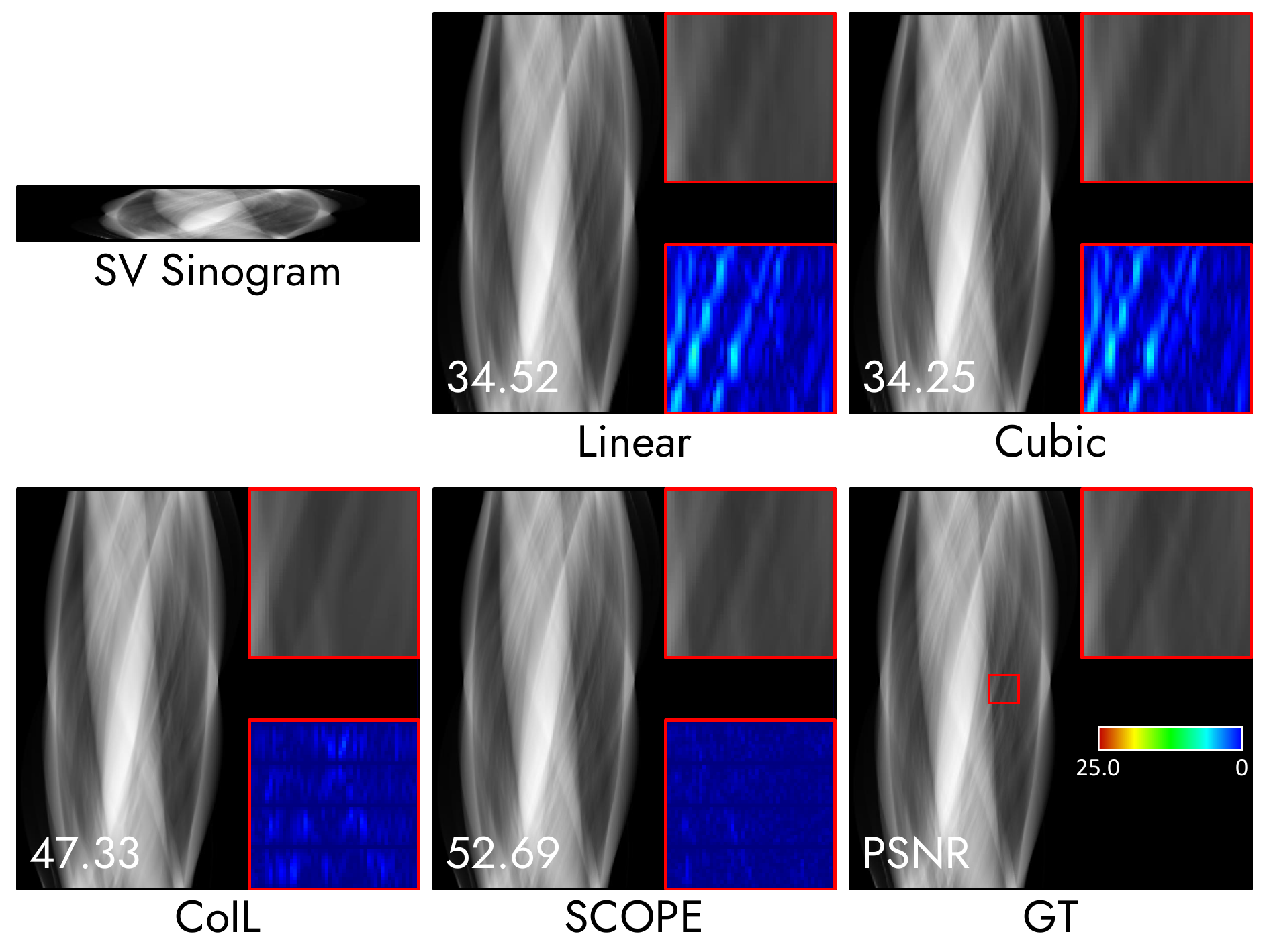}}
    \caption{Quantitative results (PSNR/SSIM) of the DV sinograms generated by different methods on a test sample ($\#90$) of the COVID-19 dataset for \textit{parallel} X-ray beam SVCT of 60 views.}
    \label{fig:sinogram}
\end{figure}
\subsubsection{Evaluation Metrics} To quantitatively measure the performance of the compared methods, we calculate Peak Signal-to-Noise Ratio (PSNR) and Structural Similarity Index Measure (SSIM) \cite{ssim}. PSNR is defined based on pixel-by-pixel distance and SSIM measures structural similarity using the mean and variance of images. 
\subsection{Effectiveness of Re-projection Reconstruction}
\label{sec:Effectiveness of Re-projection Reconstruction}
\par To validate the effectiveness of the re-projection strategy, based on the COVID-19 dataset \cite{shakouri2021covid19}, we conduct three studies for the re-projection strategy: 1) Evaluation of DV sinograms; 2) Re-projection views, \ie the number of projections in the generated DV sinograms; 3) Downstream reconstruction methods for recovering the final CT images from the DV sinograms.
\begin{figure}[t]
    \centering
    \includegraphics[width=\linewidth]{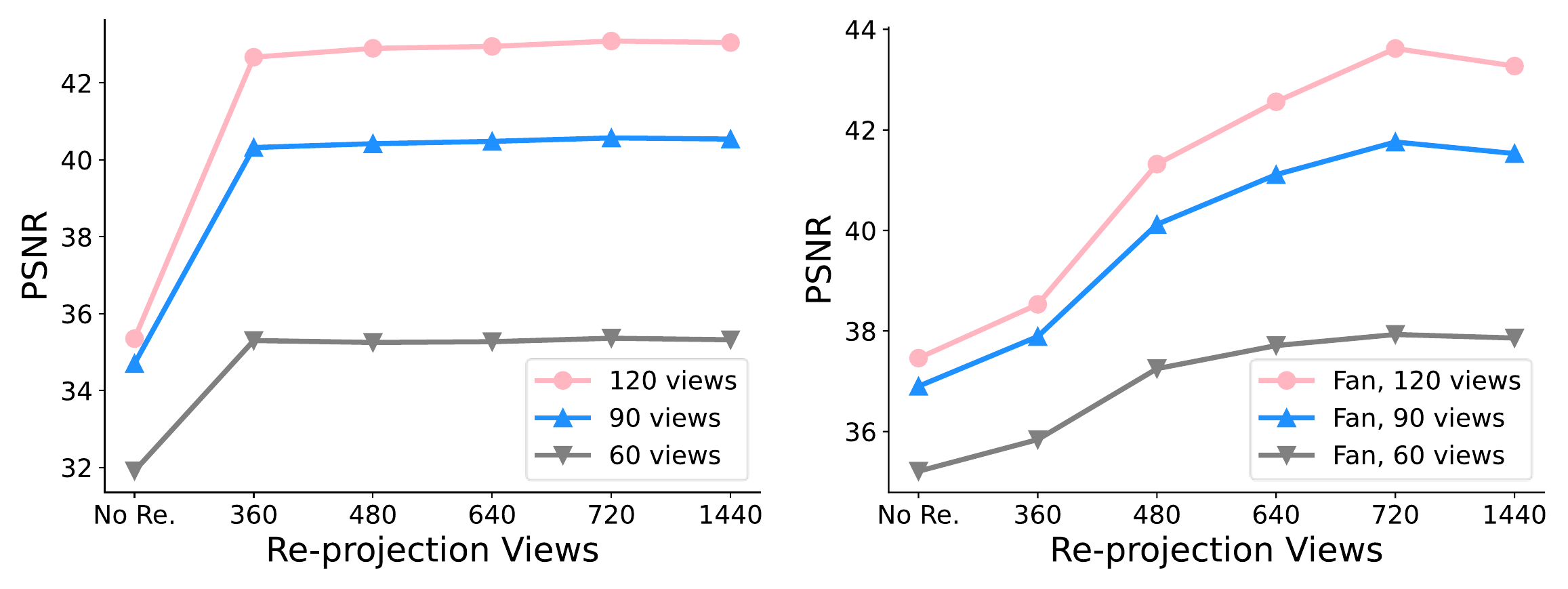}
    \caption{Quantitative results of the SCOPE model with two reconstruction strategies (no re-projection and re-projection of different numbers of projection views $K_d$) on the COVID-19 dataset for \textit{parallel} (left) and \textit{fan} (right) X-ray beam SVCT of 60, 90, and 120 views.}
    \label{fig:reprojection_curve}
\end{figure}
\subsubsection{Evaluation of DV sinograms}
\par We evaluate the performance of our SCOPE model and three other methods for generating the DV sinograms: 1) Linear interpolation; 2) Cubic interpolation; 3) CoIL \cite{sun2021coil}.
\par Table \ref{table:ablation-sinogram} shows the quantitative performance referred to the GT DV sinogram. The results show that the SCOPE model outperforms the three baselines in all cases, with PSNR improvements of 18.63 dB (53.27 vs. 34.64), 18.89 dB (53.27 vs. 34.38), and 5.98 dB (53.27 vs. 47.29) respectively, comparing with Linear interpolation in different number of views as input. Qualitative comparisons of the sinograms are shown in Fig. \ref{fig:sinogram}, where it can be observed that the DV sinogram generated by the SCOPE model exhibits clear global structures and local details, and is the closest to the ground truth sinogram.
\subsubsection{Re-projection Views}
\par We compare the following two strategies to recover the final CT image: 1) No Re-projection, we feed all the coordinates into the MLP to produce the corresponding image intensities; 2) $K_d$ Re-projection Views, we use the MLP to generate the different views of DV sinograms (360, 480, 640, 720, and 1440 views) and then apply FBP \cite{fbp} to recover the CT images.
\par
\begin{figure}[t]
    \centering
     {\includegraphics[width=0.48\textwidth]{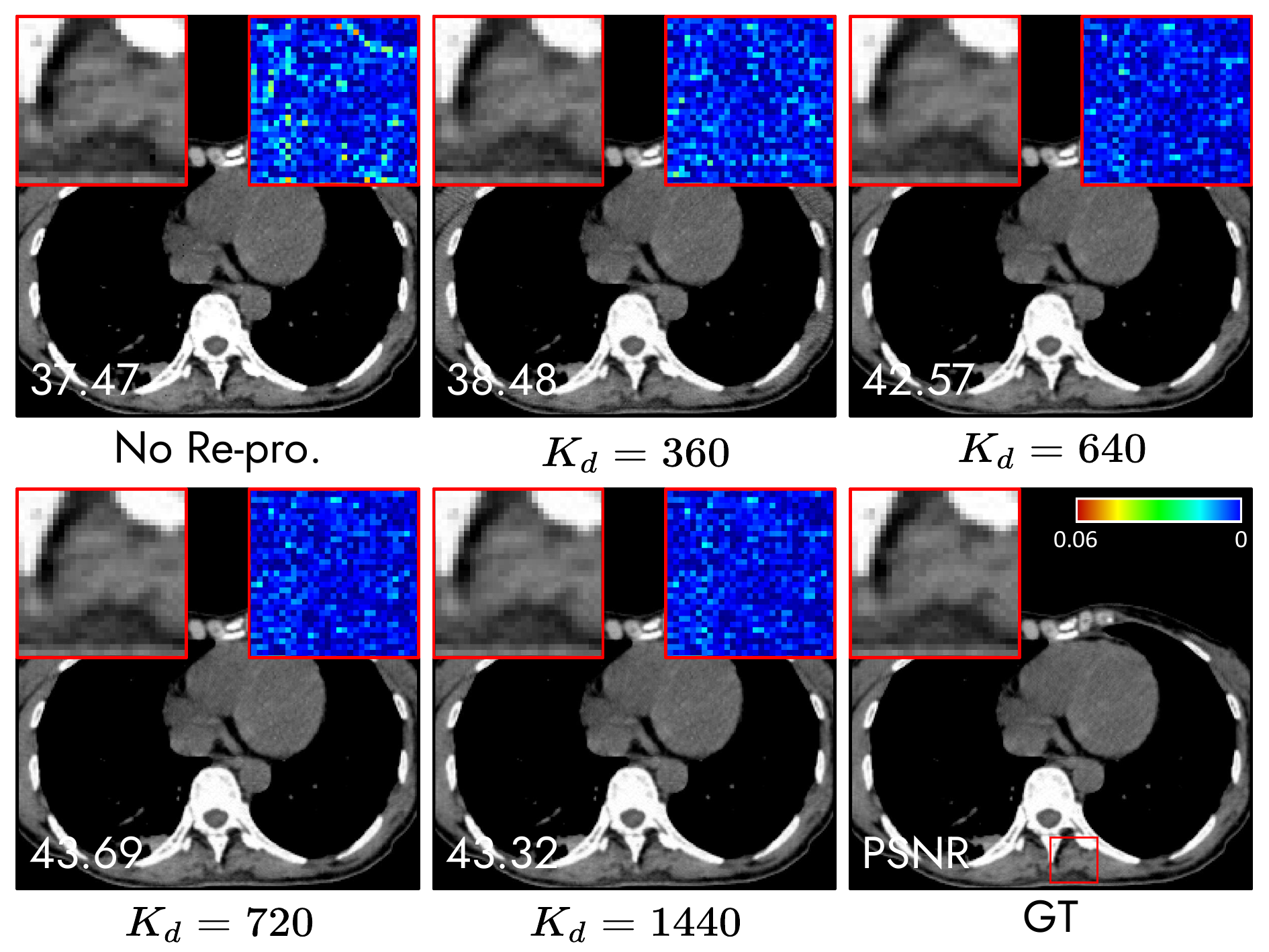}}
    \caption{Qualitative results of the SCOPE model with two reconstruction strategies (no re-projection and re-projection of different number of projection views $K_d$) on a test sample ($\#80$) of the COVID-19 dataset for \textit{fan} X-ray beam SVCT of 120 views.}
    \label{fig:reprojection}
\end{figure}

 Fig. \ref{fig:reprojection_curve} shows the quantitative results. Overall, the re-projection strategy significantly improves performance for all the cases. For example, PSNR improves by about 3 dB for fan X-ray beam SVCT of 60 views. More importantly, there is a common trend in all the cases: The model performance gradually increases when the re-projection views increase from 36w0 to 720 but slightly decreases when the re-projection views increase from 720 to 1440. Our explanation is: 1) The projections of views less than 720 are not dense enough. Although the intensity mutations of the highest frequency are completely removed, the image details of the sub-high frequency are also partially lost; 2) The projections of views more than 720 are over-dense, which results in incomplete removal of the intensity mutations and thus obtains the sub-optimal performance. Therefore, the re-projection view $K_d$ is set as 720 in this work, but it may need to be adjusted for specific cases. Fig. \ref{fig:reprojection} shows the qualitative results. The image by the direct reconstruction (\ie No. Re-pro.) contains a lot of the intensity mutations caused by the overfitting problem, while the resulting image includes some streaking artifacts due to the under-sampling when $K_d$ is 360. The results are very clear and close to the GT image when $K_d$ increases up to 640.
\subsubsection{Downstream Reconstruction Method}
\par We employ three reconstruction methods to generate the final CT images from the synthesized DV sinograms: 1) SIRT \cite{trampert1990simultaneous}, a simultaneous iterative reconstruction method; 2) FBPConvNet \cite{FBPConvNet}, a supervised DL model; 3) FBP \cite{fbp}, an analytical reconstruction algorithm. FBPConvNet is trained on the AAPM dataset.
\begin{figure}[t]
    \centering
     {\includegraphics[width=0.7\linewidth]{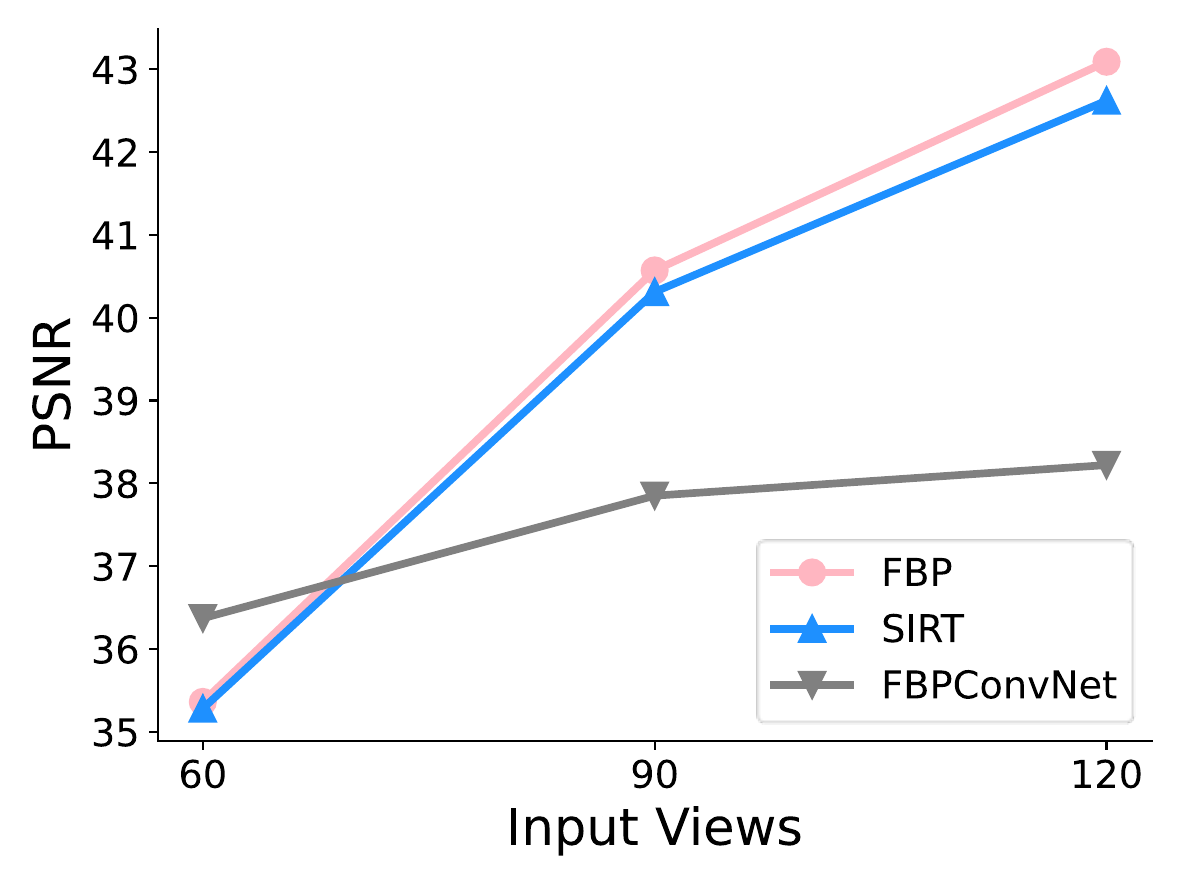}}
    \caption{Quantitative results of the SCOPE model with three downstream reconstruction methods on the COVID-19 dataset for \textit{parallel} X-ray beam SVCT of 60, 90, and 120 views.}
    \label{table:recon_method}
\end{figure}
\begin{figure}[t]
    \centering
     {\includegraphics[width=0.48\textwidth]{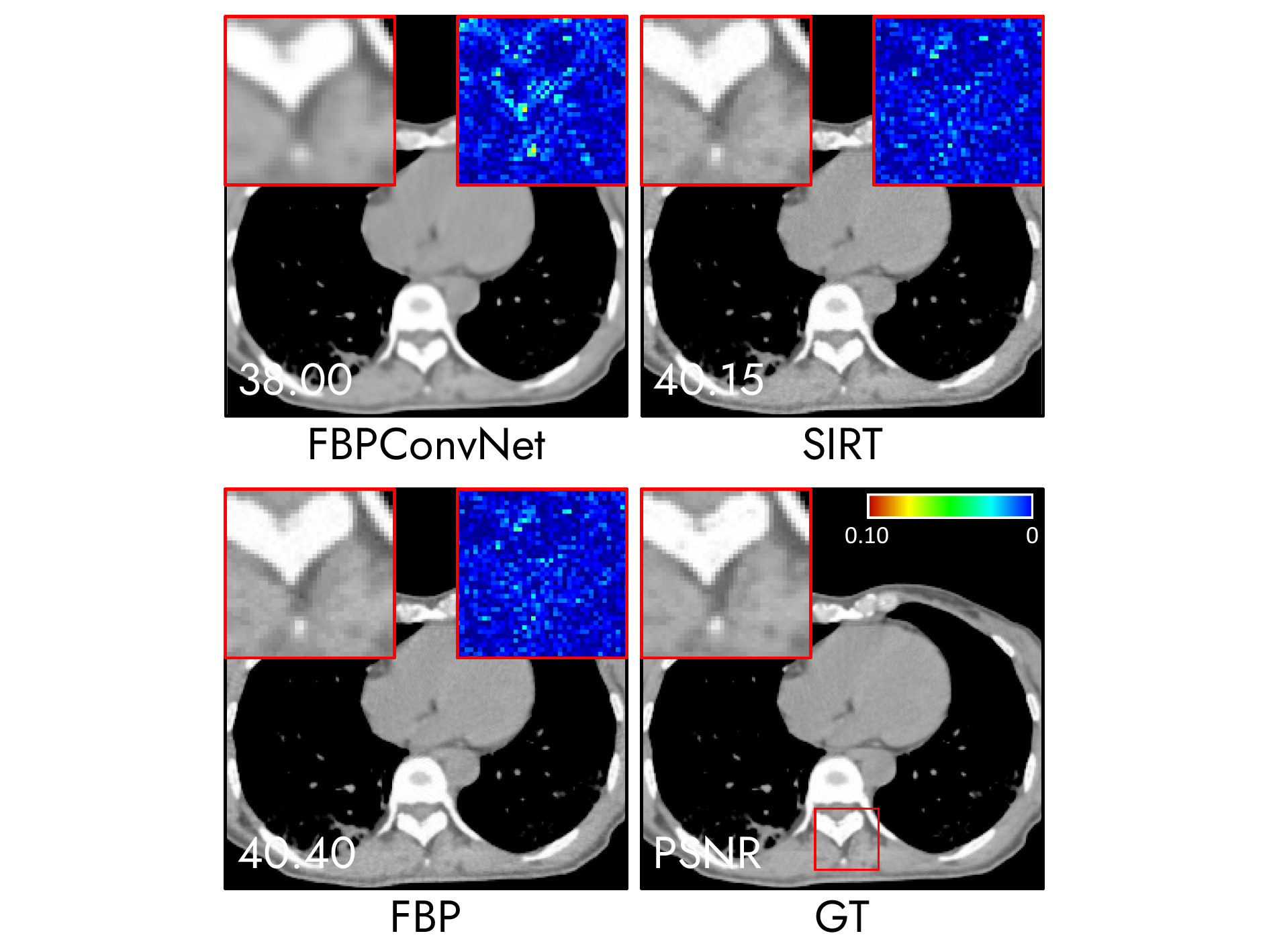}}
    \caption{Qualitative results of the SCOPE model with three downstream reconstruction methods on a test sample ($\#88$) of the COVID-19 dataset for \textit{parallel} X-ray beam SVCT of 90 views.}
    \label{fig:recon_method}
\end{figure}
\par The quantitative results are presented in Fig. \ref{table:recon_method}. It can be observed that both FBP and SIRT exhibit good performance for input views in all cases. This can be attributed to two factors. Firstly, when the input sinogram is densely sampled, FBP and SIRT tend to produce similar performances due to the stable constraint of the CT solution space. Secondly, the sinograms generated by the SCOPE model closely approximate the ground truth sinograms.
\par In addition, the FBPConvNet demonstrates the best performance for 60 input views, but the worst performance for 90 and 120 input views. The FBPConvNet is an end-to-end CNN-based denoiser that takes low-quality CT images generated from SV sinograms using FBP as inputs and outputs high-quality CT images. In our study, the inputs to FBPConvNet are the CT images obtained from DV sinograms, which are the results of the SCOPE model and FBP. The lower input view CT images with 60 views may match the FBPConvNet's training scenario, resulting in better performance than FBP and SIRT. However, for the CT images with relatively higher quality with 90 and 120 input views, FBPConvNet instead results in a loss of image details in the input CT images, thus degrading the model's performance. Fig. \ref{fig:recon_method} shows the qualitative comparison of the three reconstruction methods. It can be observed that SIRT and FBP produce CT images that are very close to the ground truth, while FBPConvNet produces an over-smoothed image.
\par To sum up, traditional reconstruction methods (including analytical and iterative methods) generally perform stable and similar on DV sinograms generated from SCOPE. Considering the computational cost, we choose FBP as the downstream reconstruction method.
\subsection{Effectiveness of Hash Encoding}
\par To validate the effectiveness of the hash encoding \cite{muller2022instant}, based on the COVID-19 dataset \cite{shakouri2021covid19}, the SCOPE models with three different encoding modules are compared: 1) No encoding, a pure 9 layers of MLP; 2) Position encoding, 9 layers of MLP with position encoding \cite{mildenhall2020nerf}; 3) Hash encoding, 3 layers of MLP with hash encoding \cite{muller2022instant}.
\begin{figure}[t]
    \centering
     {\includegraphics[width=0.48\textwidth]{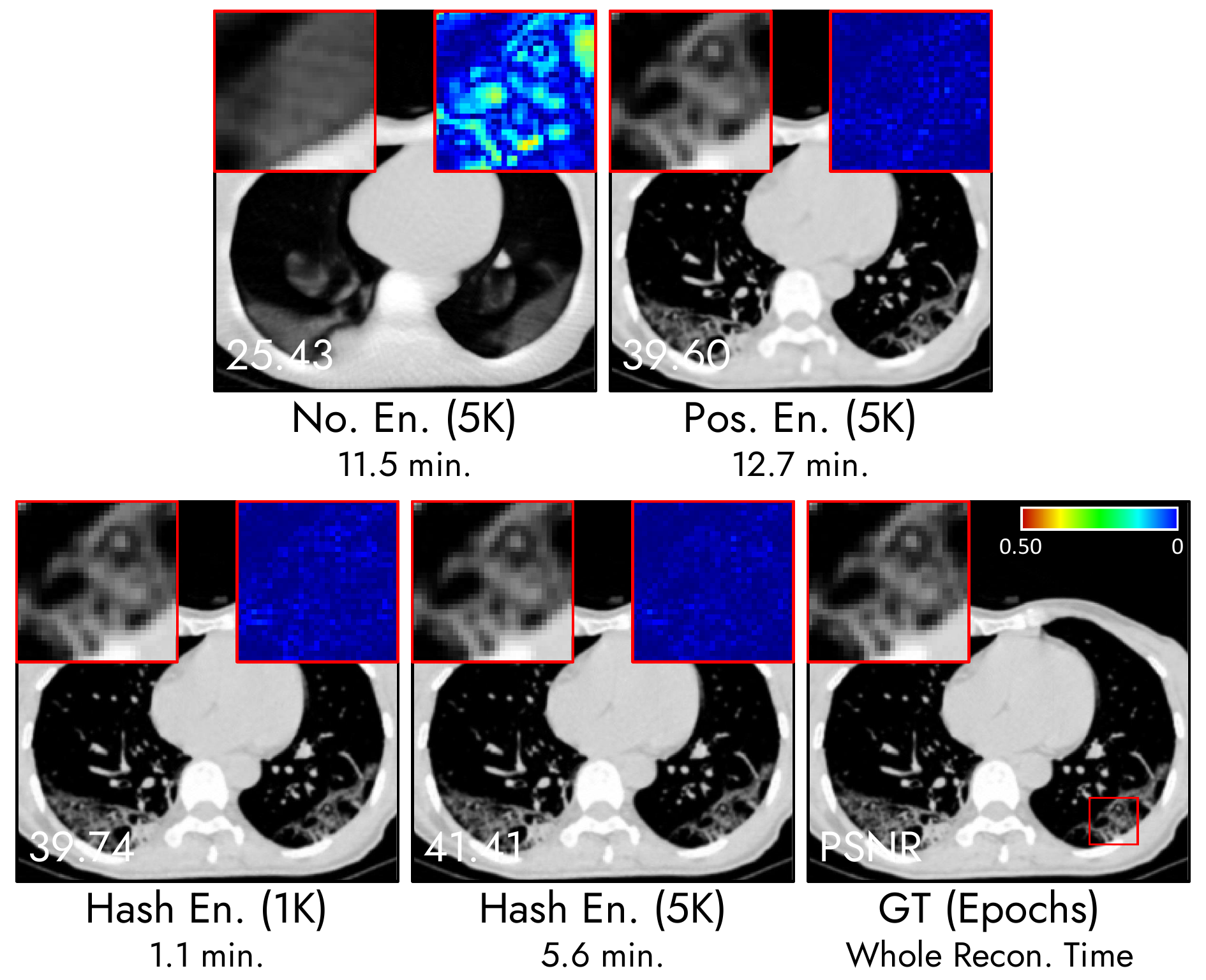}}
    \caption{Qualitative results of the SCOPE model with three encoding modules on a test sample ($\#95$) of the COVID-19 dataset for \textit{fan} X-ray beam SVCT of 90 views.}
    \label{fig:encoding_fan}
\end{figure}
\begin{table}[t]
\centering
\caption{Quantitative results (PSNR/SSIM) of the SCOPE model with three encoding modules on the COVID-19 dataset for \textit{parallel} and \textit{fan} X-rays beam SVCT of 60, 90, and 120 views. The best performances are highlighted in bold.}
\label{table:ablation-encoding}
\resizebox{0.93\linewidth}{!}{
\begin{tabular}{clccc}
\toprule
\textbf{X-ray}                & \textbf{Encoding} & \textbf{60 Views} & \textbf{90 Views} & \textbf{120 Views}  \\ 
\midrule
\multirow{3}{*}{\textit{Parallel}} & No En.            & $27.52/0.8378$      & $28.48/0.8632$      & $29.30/0.8831$        \\
                                   & Pos. En.          & $35.19/0.8870$      & $38.65/0.9207$      & $40.05/0.9294$        \\
                                   & Hash En.          & $\mathbf{35.36/0.9512}$      & $\mathbf{40.57/0.9807}$      & $\mathbf{43.09/0.9872}$        \\ 
\midrule
\multirow{3}{*}{\textit{Fan}}      & No En.            & $24.10/0.7256$      & $25.97/0.7828$      & $26.51/0.8017$        \\
                                   & Pos. En.          & $36.27/0.9539$      & $39.56/0.9760$      & $41.73/0.9839$        \\
                                   & Hash En.          & $\mathbf{37.93/0.9727}$      & $\mathbf{41.76/0.9854}$      & $\mathbf{43.62/0.9888}$        \\
\bottomrule
\end{tabular}}
\end{table}
\begin{figure}[t]
    \centering
    \includegraphics[width=0.7\linewidth]{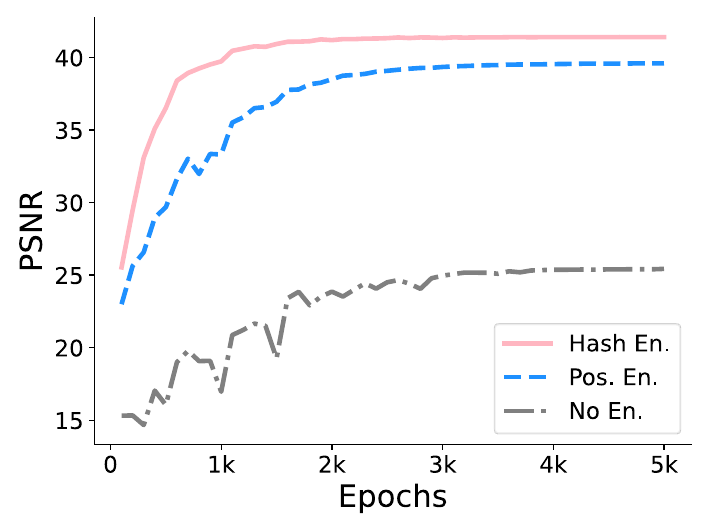}
    \caption{Performance curves of the SCOPE model with three encoding modules over training epochs on a test sample ($\#95$) of the COVID-19 dataset for \textit{fan} X-ray beam SVCT of 90 views.}
    \label{fig:encoding-curves}
\end{figure}
\par Table \ref{table:ablation-encoding} shows the quantitative results. From the results, we see that, compared with no encoding, both the position encoding and hash encoding significantly improve the model performance in terms of all three metrics for all the cases. For example, PSNR respectively improves by 13.59 dB (39.56 vs. 25.97) and 15.79 dB (41.76 vs. 25.97) for fan X-ray beam SVCT of 90 views. This is due to the spectral bias problem \cite{rahaman2019spectral, xu2019frequency}, \ie a pure MLP is biased toward learning low-frequency signals during the practical training. Therefore, encoding modules are critical for improving the MLP's ability to learn high-frequency signals. Besides, we observe that the hash encoding slightly outperforms the position encoding in most cases. E.g., PSNR improves by 1.66 dB (37.93 vs. 36.27) for fan X-ray beam SVCT of 60 views. Fig. \ref{fig:encoding_fan} shows the qualitative results on a test sample ($\#$95) for fan X-ray beam SVCT of 90 views. Overall, hash encoding achieves the best image quality and the fastest reconstruction speed benefiting from its adaptive encoding and the shallower MLP. Numerically, hash encoding takes only about 1.1 $\mathsf{min}$ to obtain the same performance as position encoding. However, position encoding takes 12.7 $\mathsf{mins}$, which is about 12$\times$ time cost. We also show the performance curves of the three encoding modules over training epochs in Fig. \ref{fig:encoding-curves}. Obviously, hash encoding produces the best performance.

\subsection{Disentangling Re-projection versus Hash Encoding}
\par To disentangle the effect of the re-projection reconstruction versus hash encoding \cite{muller2022instant} for the resulting image quality, we train two versions of SCOPE models (positional encoding \cite{mildenhall2020nerf} and hash encoding) on the COVID-19 dataset \cite{shakouri2021covid19}. Then, based on the well-trained models, we use two reconstruction strategies (w/o and w/ re-projection strategy) to generate the final CT images.
\par
\begin{figure}[t]
    \centering
    \includegraphics[width=0.48\textwidth]{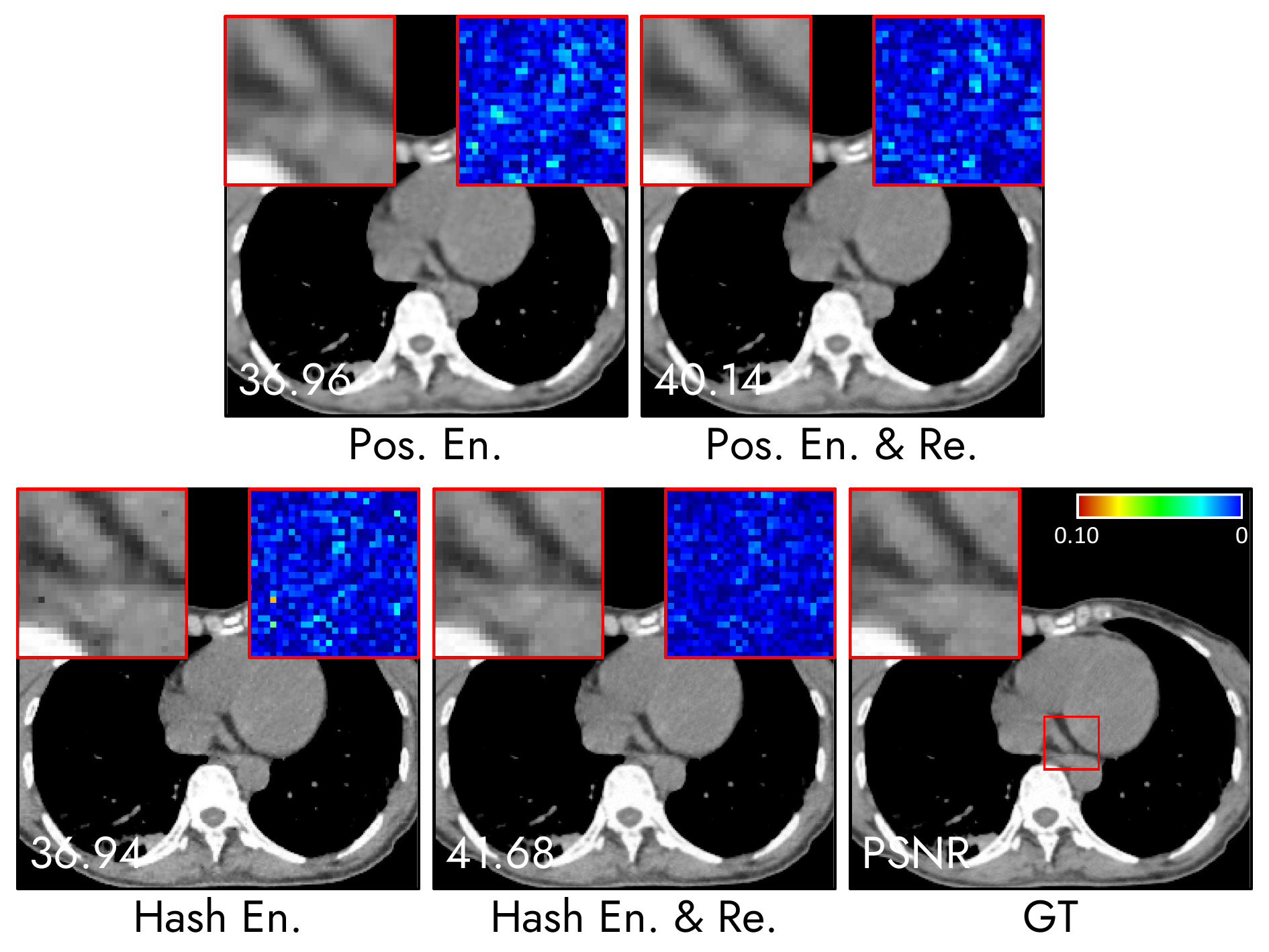}
    \caption{Qualitative results of four versions of the SCOPE models on a test sample ($\#80$) of the COVID-19 dataset for \textit{fan} X-ray beam SVCT of 90 views.}
    \label{fig:fig_disen_encoding_reprojection}
\end{figure}

 Fig. \ref{fig:fig_disen_encoding_reprojection} shows the qualitative results. It is shown that without the re-projection, hash encoding produces a sharper CT reconstruction image than positional encoding. The residual map of positional encoding illustrates image detail loss along tissue edges, which is not present in hash encoding. This is consistent with previous studies \cite{muller2022instant, tewari2022advances} indicating that hash encoding enables more efficient learning for high-frequency image content due to its adaptive encoding strategy. However, the residual map of hash encoding indicates a few noise-like image intensity mutations, which might be caused by overfitting to the noise in the SV sinogram. When referring to the quantitative comparison in Table \ref{table:dis_enc_repro_table}, it is notable that the image quality is not improved based on hash encoding. For example, hash encoding only enables 0.27 dB improvement on PSNR from positional encoding in 90 views, and reduces 0.25 dB in 120 views. However, after performing the reprojection process, both cases exhibit remarkable improvement in terms of image quality and quantitative evaluations. The image mutations are effectively suppressed in residual maps, and the PSNR results are significantly increased. For instance, the re-projection method contributes a 4.86 dB improvement in PSNR compared to hash encoding in 90 views.
\begin{table}
\centering
\caption{Quantitative results of four versions of the SCOPE models on the COVID-19 dataset for \textit{fan} X-ray beam SVCT of 60, 90, and 120 views. The best performances are highlighted in bold.}
\label{table:dis_enc_repro_table}
\resizebox{0.465\textwidth}{!}{
\begin{tabular}{lccc} 
\toprule
\textbf{Modules} & \textbf{60 views} & \textbf{90 views} & \textbf{120 views}  \\ 
\midrule
Pos.En.           & $34.62/0.9439$      & $36.63/0.9661$      & $37.71/0.9738$        \\
Hash En.          & $35.21/0.9616$      & $36.90/0.9738$      & $37.46/0.9769$        \\
Pos.En. \& Re.  & $36.27/0.9539$      & $39.56/0.9760$      & $41.73/0.9839$        \\
Hash En. \& Re. & $\mathbf{37.93/0.9727}$      & $\mathbf{41.76/0.9854}$      & $\mathbf{43.62/0.9888}$        \\
\bottomrule
\end{tabular}}
\end{table}
\par In summary, both the qualitative and quantitative results indicate that compared with hash encoding, our re-projection reconstruction strategy contributes more to the reconstructed image quality.
\subsection{Influence of Noises}
\par During the CT acquisition, noises are inevitable due to various factors (\eg background noises). Based on the SV sinograms $\mathbf{y}_s$ of 90 views from the COVID-19 dataset \cite{shakouri2021covid19}, we simulate the measurement domain noise via a statistical Poisson noise model \cite{elbakri2002statistical, yin2019domain}, which is defined as:
\begin{linenomath*}
\begin{equation}
    \mathbf{Y}(\theta, \rho)\sim \text{Poisson}\left\{b\times e^{-\mathbf{y}_s(\theta, \rho)}+r\right\},
\end{equation}
\end{linenomath*}
where $\mathbf{Y}(\theta, \rho)$ denotes the transmitted X-ray photon intensity, $b$ is the incident X-ray photon intensity, and $r$ is the mean of the background events and read-out noise variance.
\par We set the value of $r$ to 10, and the value of $b$ to $1.3\times 10^4$, $4\times10^4$, and $4\times10^5$ to simulate three levels of noise (SNR $\approx$ 35 dB, 40 dB, and 50 dB) respectively. We then generate noisy sinograms $\mathbf{y}_{s}^{\prime}(\theta, \rho)=-\ln\left({\mathbf{Y}\left(\theta, \rho\right)}/{b}\right)$ using a negative logarithm. Two versions of the SCOPE model (with and without re-projection) and five baseline models are compared for recovering the CT images from the noisy sinograms.
\begin{table}
\centering
\caption{uantitative results (PSNR/SSIM) of different methods on the COVID-19 dataset for \textit{fan} X-rays beam SVCT of 90 views with three levels of noises (SNR = 35 dB, 40 dB, and 50 dB). The best performances are highlighted in bold.}
\label{table:noise_table}
\resizebox{0.47\textwidth}{!}{
\begin{tabular}{lccc} 
\toprule
\textbf{Method} & \textbf{35 dB} & \textbf{40 dB} & \textbf{50 dB}  \\ 
\midrule
FBP \cite{fbp}            & $15.90/0.2452$   & $17.68/0.3018$   & $18.73/0.3539$    \\
CoIL \cite{sun2021coil}           & $25.88/0.5919$   & $27.59/0.6757$   & $28.52/0.7284$    \\
GRFF \cite{tancik2020fourier}           & $\mathbf{30.48/0.8825}$   & $32.63/\mathbf{0.9177}$   & $35.46/0.9529$    \\
SCOPE (w/o Re.)  & $27.43/0.8282$   & $30.33/0.9071$   & $34.54/0.9664$    \\
SCOPE (w/ Re.)   & $29.94/0.8342$   & $\mathbf{32.76}/0.9172$   & $\mathbf{36.85/0.9728}$    \\
\bottomrule
\end{tabular}}
\end{table}
\begin{figure}[t]
    \centering
    \includegraphics[width=0.48\textwidth]{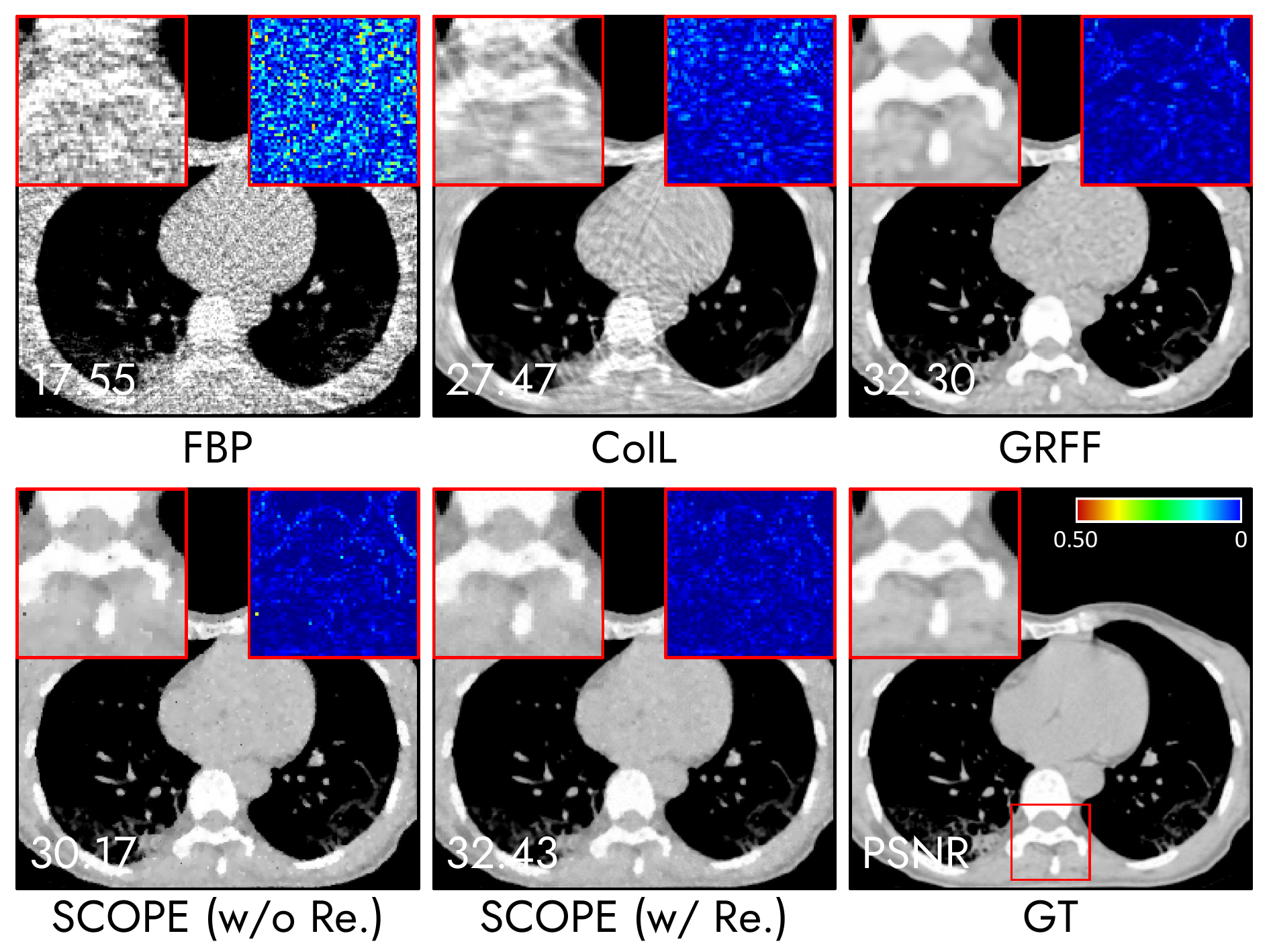}
    \caption{Qualitative results of different methods on a test sample ($\#95$) of the COVID-19 dataset for \textit{fan} X-rays beam SVCT of 90 views with the noises of SNR $\approx$ 40 dB.}
    \label{fig:fig_noise}
\end{figure}
\begin{table}[ht]
\centering
\caption{Quantitative results (PSNR/SSIM) of all the compared methods on the AAPM and COVID-19 datasets for \textit{parallel} X-rays beam SVCT of 60, 90, and 120 views. The best performances are highlighted in bold.}
\label{table:ablation-compraison_parallel}
\resizebox{\linewidth}{!}{
\begin{tabular}{clcccc} 
\toprule
\textbf{Dataset}                   & \textbf{Method} & \textbf{60 Views} & \textbf{90 Views} & \textbf{120 Views}  \\ 
\midrule
\multirow{6}{*}{\textbf{AAPM}}     & FBP \cite{fbp}            & $19.98/0.2791$      & $24.40/0.4328$      & $28.30/0.5869$        \\
                                   & CoIL \cite{sun2021coil}           & $32.35/0.7548$      & $36.52/0.8685$      & $39.11/0.9276$        \\
                                   & GRFF \cite{tancik2020fourier}           & $35.99/0.9448$      & $38.89/0.9664$      & $39.71/0.9716$        \\
                                   & FBPConvNet \cite{FBPConvNet}     & $38.66/0.9392$      & $41.95/0.9557$      & $44.14/0.9644$        \\
                                   & TF U-Net \cite{han2018framing}       & $\mathbf{38.67}/0.9388$      & $41.74/0.9587$      & $43.86/0.9688$        \\
                                   & SCOPE           & $38.05/\mathbf{0.9596}$      & $\mathbf{42.18/0.9794}$      & $\mathbf{44.33/0.9860}$        \\ 
\midrule
\multirow{6}{*}{\textbf{COVID-19}} & FBP \cite{fbp}            & $22.53/0.4979$      & $27.56/0.6722$      & $31.60/0.8025$        \\
                                   & CoIL \cite{sun2021coil}           & $30.12/0.8075$      & $34.54/0.9143$      & $37.71/0.9519$        \\
                                   & GRFF \cite{tancik2020fourier}           & $33.76/0.9494$      & $35.51/0.9676$      & $35.95/0.9702$        \\
                                   & FBPConvNet \cite{FBPConvNet}     & $32.87/0.9236$      & $36.65/0.9563$      & $38.40/0.9661$        \\
                                   & TF U-Net \cite{han2018framing}        & $32.86/0.9297$      & $36.59/0.9571$      & $38.71/0.9675$        \\
                                   & SCOPE           & $\mathbf{35.36/0.9512}$      & $\mathbf{40.57/0.9807}$      & $\mathbf{43.09/0.9872}$        \\
\bottomrule
\end{tabular}}
\end{table}
\par Table \ref{table:noise_table} shows the quantitative results. There are some observations: 1) As the noise level increases, the performance of all models decreases, which is expected. Since the noise result in the SVCT inverse imaging problem being more ill-posed and thus the models may produce local minimal solutions; 2) The SCOPE model achieves the best performance at the low noise level (50 dB SNR), while the GRFF model performs best at the high noise level (35 dB SNR). Compared with Fourier encoding in GRFF, hash encoding in SCOPE tend to fit to high-frequency image content more efficiently, while may cause overfitting to noise; 3) Our re-projection strategy remarkably improves the model performance. Fig. \ref{fig:fig_noise} illustrates the qualitative results. SCOPE model with the re-projection process produces the least error in residual map.
\par To sum up, the performance of all the compared models is limited by noise in the measurement domain and our SCOPE model achieves the best performance in most cases.
\begin{figure*}[t]
    \centering
     {\includegraphics[width=0.9\linewidth]{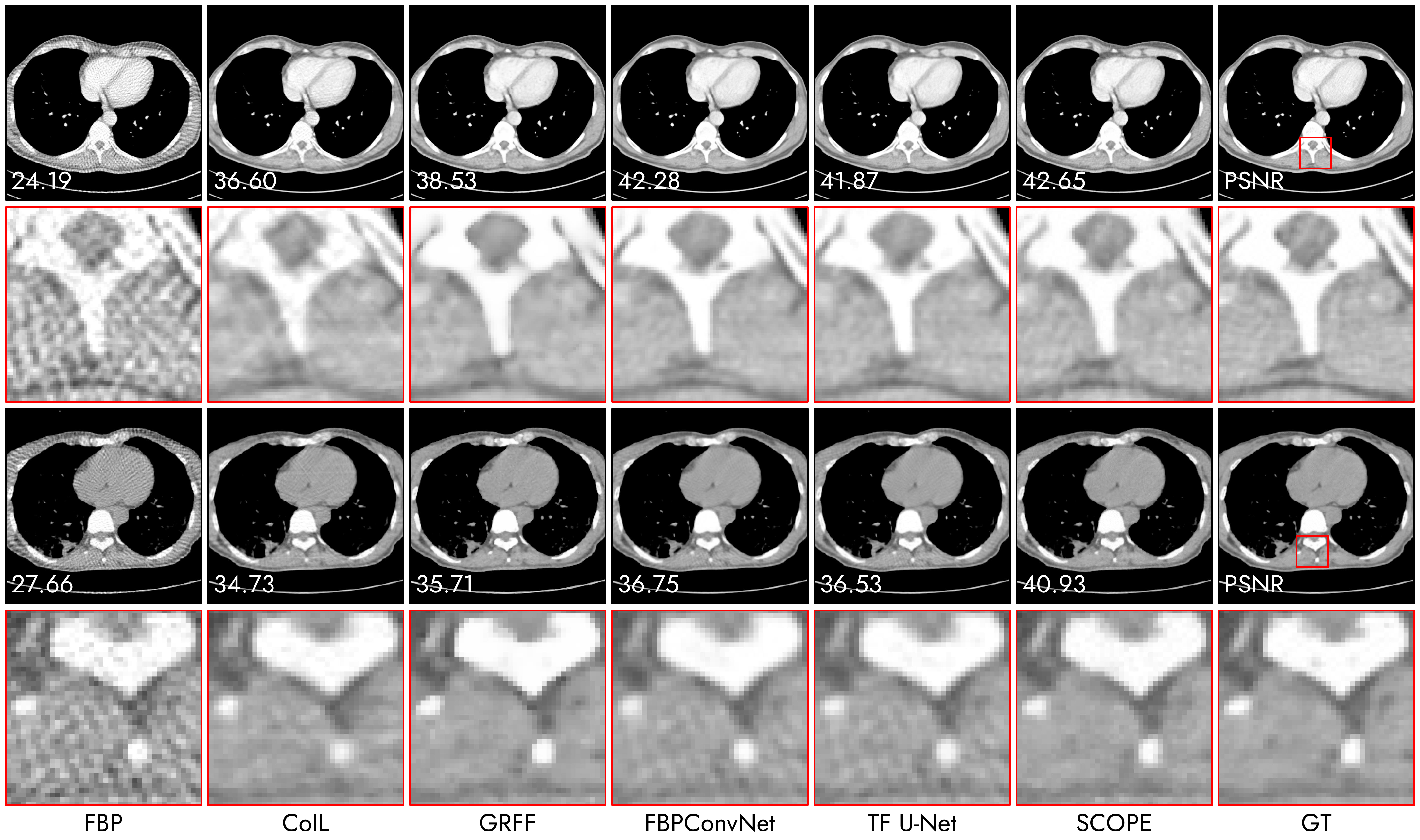}}
    \caption{Qualitative results of all the compared methods on two test samples ($\#$109 (Rows 1-2) and $\#$90 (Rows 3-4)) of the AAPM dataset and COVID-19 dataset for \textit{parallel} X-ray beam SVCT of 90 views.}
    \label{fig:com_parallel}
\end{figure*}
\begin{figure*}[t]
    \centering
     {\includegraphics[width=0.9\linewidth]{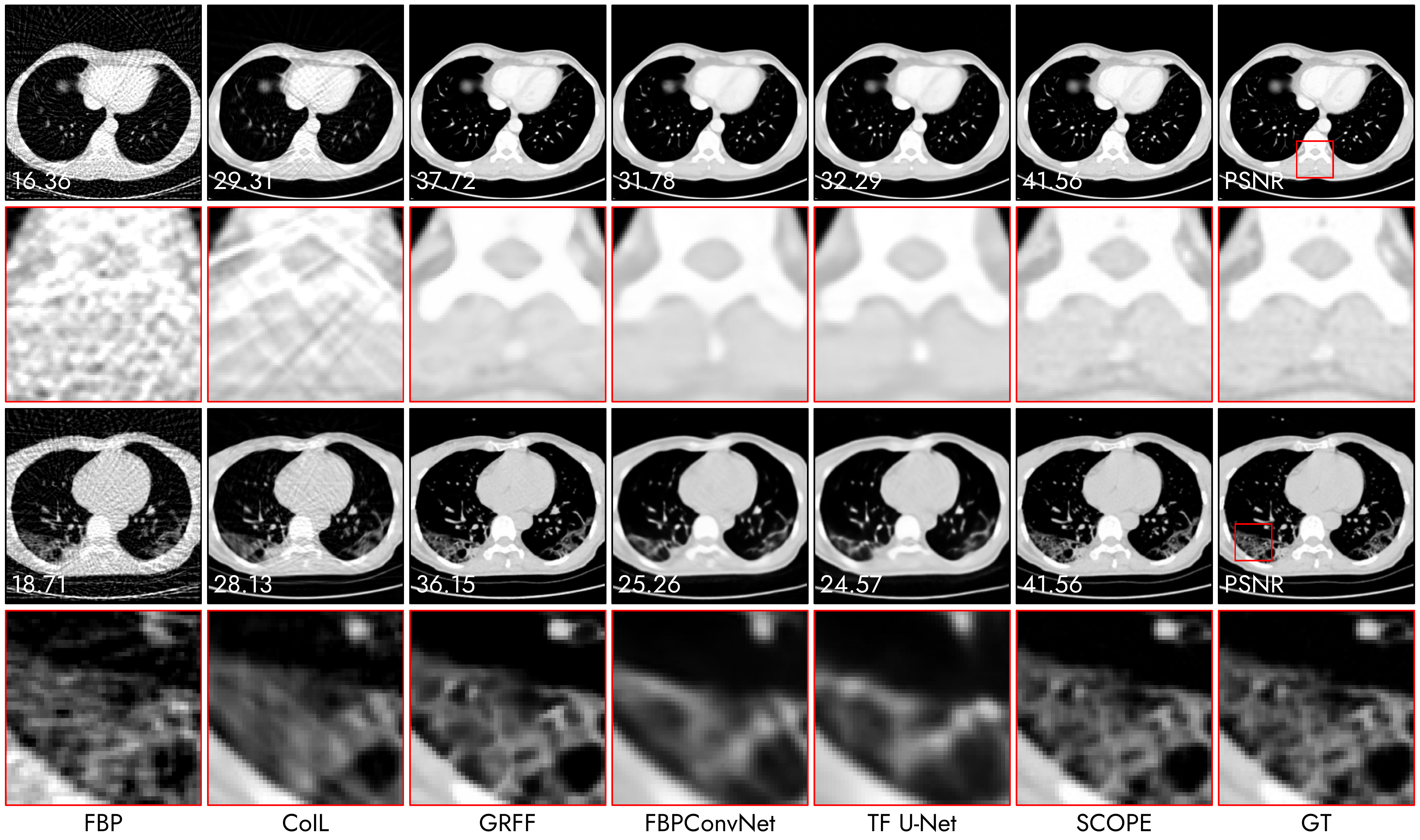}}
    \caption{Qualitative results of all the compared methods on two test samples ($\#$104 (Rows 1-2) and $\#$95 (Rows 3-4)) of the AAPM dataset and COVID-19 dataset for \textit{fan} X-ray beam SVCT of 90 views.}
    \label{fig:com_fan}
\end{figure*}
\subsection{Comparison with Other Methods}
\par Finally, we compare the proposed SCOPE model with the five baselines on the AAPM and COVID-19 datasets for parallel and fan X-ray beam SVCT reconstruction of 60, 90, and 120 input views. Since FBPConvNet \cite{FBPConvNet} and TF U-Net\cite{han2018framing} are supervised DL methods, we train them on the training set of the AAPM dataset. Other four methods (FBP \cite{fbp}, CoIL \cite{sun2021coil}, GRFF \cite{tancik2020fourier}, and our SCOPE model) are image-specific and thus they direct reconstruct the corresponding high-quality CT image from each SV sinogram. Note that the parallel and fan X-ray beam SVCT are considered two independent reconstruction problems and thus all the training and test processes are solely conducted.
\subsubsection{Parallel X-ray Beam SVCT} Table \ref{table:ablation-compraison_parallel} shows the quantitative results. On the AAPM dataset, the SCOPE produces the best performance for most cases. Compared with the two supervised DL methods (FBPConvNet \cite{FBPConvNet} and TF U-Net \cite{han2018framing}), the SCOPE also obtains minor performance improvements. For instance, PSNR respectively improves by 0.27 dB (42.18 vs. 41.95) and 0.44 dB (42.18 vs. 41.74) when 90 input views. On the COVID-19 dataset, we, however, observe that FBPConvNet and TF U-Net  suffer from severe performance drops. This is mainly due to the domain shift problem, \ie the training and test data do not share the same distribution. In comparison, the SCOPE model still produces excellent reconstruction results on the COVID-19 data because it is image-specific. For example, the difference in PSNR between SCOPE and FBPConvNet is up to +3.92 dB (40.57 vs 36.65) when 90 input views. Fig. \ref{fig:com_parallel} shows the qualitative results on two test samples ($\#109$ and $\#90$) of the AAPM and COVID-19 datasets. On the test sample $\#109$ from the AAPM dataset, both FBP \cite{fbp} and CoIL \cite{sun2021coil} can not produce satisfactory results, which still include a lot of streaking artifacts. GRFF \cite{tancik2020fourier} yields a smooth result that lost some image details. In comparison, FBPConvNet, TF U-Net, and SCOPE all recover the desirable images that are hardly distinguished from GT images. On the test sample $\#90$ from the COVID-19 dataset, the two supervised models obtain sub-optimal results including moderate streaking artifacts, while our SCOPE model still produces a high-quality image that is closest to the GT image.
\subsubsection{Fan X-ray Beam SVCT} Table \ref{table:ablation-compraison_fan} shows the quantitative results. We observe that the SCOPE and GRFF \cite{tancik2020fourier} respectively produce the best and second-best performance in terms of all three metrics for all the cases. For example, on the AAPM dataset for 90 input views, SCOPE and GRFF respectively achieve 40.92 dB and 37.54 dB, while TF U-Net \cite{han2018framing} only obtains 32.47 dB in terms of PSNR. It is not common that FBPConvNet \cite{FBPConvNet} and TF U-Net cannot produce a satisfactory performance on the AAPM dataset although they are trained on the AAPM dataset. We believe that, for learning the end-to-end mapping as in the supervised DL methods, the fan X-ray beam SVCT is a more difficult task than the parallel X-ray beam SVCT when the same input views. In our experiments, given the sinograms of the same projection views, the results of the fan X-ray CT include more severe streaking artifacts than that of the parallel X-ray CT after applying the FBP algorithm \cite{fbp}. While FBPConvNet \cite{FBPConvNet} and TF U-Net \cite{han2018framing} directly learn the inverse mapping from the artifacts-corrupted inputs to the artifacts-free outputs. Therefore, they are not expected to perform as well in the fan X-ray CT as in the parallel X-ray CT. In contrast, GRFF \cite{tancik2020fourier} and SCOPE train MLP networks to learn the implicit function of the unknown CT image by computing the loss on the SV sinogram (\ie they do not manipulate image information directly). Thus, they all work well for different types of X-ray beams. Fig. \ref{fig:com_parallel} shows the qualitative results on two test samples ($\#104$ and $\#95$) of the AAPM and COVID-19 datasets. We see that the four compared methods do not recover good results. The results from FBP algorithm \cite{fbp} and CoIL \cite{sun2021coil} include severe streaking artifacts, while FBPConvNet \cite{FBPConvNet} and TF U-Net \cite{han2018framing} produce the overly smooth results. GRFF \cite{tancik2020fourier} obtains the second-best results that lost some image details. Only the proposed SCOPE removes streaking artifacts greatly and preserves fine image details well.

\section{Conclusion}
\label{sec:dis_conc}
\par In this work, we propose SCOPE, a self-supervised INR-based method for SVCT reconstruction. Like previous INR works \cite{tancik2020fourier,shen2021nerp,zang2021intratomo}, SCOPE represents the desired CT image as an implicit continuous function and trains a neural network to learn the implicit function by minimizing predicted errors on the acquired SV sinogram. Benefiting from image continuity prior imposed by the implicit function and neural network architecture prior, the function can be estimated. However, the solution is not optimal due to the overfitting problem. To this end, we propose a simple and effective re-projection strategy that greatly improves the resulting CT image quality. Besides, we adopt the recent hash encoding \cite{muller2022instant} into our SCOPE to accelerate the model training greatly. Experimental results on two publicly available datasets indicate that the proposed SCOPE model is not only superior to two last INR-based methods, but also outperforms two well-known supervised CNN-based methods, qualitatively and quantitatively.
\begin{table}[ht]
\centering
\caption{Quantitative results (PSNR/SSIM) of all the compared methods on the AAPM and COVID-19 datasets for \textit{fan} X-rays beam SVCT of 60, 90, and 120 views. The best performances are highlighted in bold.}
\label{table:ablation-compraison_fan}
\resizebox{\linewidth}{!}{
\begin{tabular}{clccc} 
\toprule
\textbf{Dataset}                   & \textbf{Method} & \textbf{60 Views} & \textbf{90 Views} & \textbf{120 Views}  \\ 
\midrule
\multirow{6}{*}{\textbf{AAPM}}     & FBP \cite{fbp}            & $13.74/0.1224$      & $16.40/0.1882$      & $19.34/0.2721$        \\
                                   & CoIL \cite{sun2021coil}           & $27.48/0.6171$      &  $30.87/0.7145$     & $33.02/0.7739$      \\
                                   & GRFF \cite{tancik2020fourier}          & $34.48/0.9268$      & $37.54/0.9533$      & $38.88/0.9621$        \\
                                   & FBPConvNet \cite{FBPConvNet}     & $28.72/0.8622$      & $31.95/0.9103$      & $34.56/0.9337$        \\
                                   & TF U-Net \cite{han2018framing}       & $28.31/0.8274$      & $32.47/0.8910$      & $34.94/0.9226$        \\
                                   & SCOPE           & $\mathbf{37.93/0.9560}$      & $\mathbf{40.92/0.9719}$      & $\textbf{42.76/0.9788}$        \\ 
\midrule
\multirow{6}{*}{\textbf{COVID-19}} & FBP \cite{fbp}            & $15.15/0.2230$      & $18.87/0.3642$      & $21.95/0.4832$        \\
                                   & CoIL \cite{sun2021coil}           &  $25.11/0.5978$      &  $28.54/0.7274$    & $31.32/0.8244$     \\
                                   & GRFF \cite{tancik2020fourier}           & $34.00/0.9401$      & $36.31/0.9607$      & $37.16/0.9679$        \\
                                   & FBPConvNet \cite{FBPConvNet}     & $23.03/0.7274$      & $25.58/0.8030$      & $27.28/0.8443$        \\
                                   & TF U-Net \cite{han2018framing}       & $23.12/0.7044$      & $25.05/0.7795$      & $26.43/0.8169$        \\
                                   & SCOPE           & $\mathbf{37.93/0.9727}$      & $\mathbf{41.76/0.9854}$      & $\mathbf{43.62/0.9888}$        \\
\bottomrule
\end{tabular}}
\end{table}

\bibliographystyle{IEEEtran}
\bibliography{ref}
\end{document}